\pdfoutput=1

\documentclass[reprint,aip,jcp,showpacs,twocolumn]{revtex4-1}
\usepackage{mathptmx}
\usepackage{epsfig,amsopn}
\usepackage{graphicx}
\usepackage{amsmath,amssymb}
\usepackage{natbib}
\usepackage{xcolor}
\usepackage{braket}
\usepackage{float}
\usepackage{enumerate}
\usepackage{mathtools}
\usepackage[symbol]{footmisc}
\usepackage{appendix}
\usepackage{stackengine}
\allowdisplaybreaks[1]

\newcommand{\eref}[1]{Eq.~(\ref{#1})}%
\newcommand{\fref}[1]{Fig.~\ref{#1}} %

\def\bea{\begin{eqnarray}}
\def\eea{\end{eqnarray}}

\def\p{\partial}

\begin{document}

\title{Diffusion with resetting in a logarithmic potential}

\author{Somrita Ray}
\email{somritaray@mail.tau.ac.il}
\author{Shlomi Reuveni}
\email{shlomire@tauex.tau.ac.il}

\affiliation{\noindent \textit{
School of Chemistry, The Center for Physics and Chemistry of Living Systems, The Raymond and Beverly Sackler Center for Computational Molecular and Materials Science, \& The Ratner Center for Single Molecule Science, Tel Aviv University, Tel Aviv 69978, Israel}}

\date{\today}
\begin{abstract}
\noindent
We study the effect of resetting on diffusion in a logarithmic potential. In this model, a particle diffusing in a potential $U(x) = U_0\log|x|$ is reset, i.e., taken back to its initial position, with a constant rate $r$. We show that this analytically tractable model system exhibits a series of phase transitions as a function of a single parameter, $\beta U_0$, the ratio of the strength of the potential to the thermal energy. For $\beta U_0<-1$ the potential is strongly repulsive, preventing the particle from reaching the origin. Resetting then generates a non-equilibrium steady state which is characterized exactly and thoroughly analyzed. In contrast, for $\beta U_0>-1$ the potential is either weakly repulsive or attractive and the diffusing particle eventually reaches the origin. In this case, we provide a closed form expression for the subsequent first-passage time distribution and show that a resetting transition occurs at $\beta U_0=5$. Namely, we find that resetting can expedite arrival to the origin when $-1<\beta U_0<5$, but not when $\beta U_0>5$. The results presented herein generalize results for simple diffusion with resetting --- a widely applicable model that is obtained from ours by setting $U_0=0$. Extending to general potential strengths, our work opens the door to theoretical and experimental investigation of a plethora of problems that bring together resetting and diffusion in logarithmic potential. 
\end{abstract}




\maketitle
\section{Introduction}
\indent
Resetting, i.e., stopping an ongoing dynamical process to start it anew, has recently gained significant attention due to its prevalence in natural and man made systems. For example, it has long been known that resetting certain computer algorithms can significantly enhance their performance \cite{CS1,CS2,CS3,CS4}. Resetting is also commonly considered in the context of search processes \cite{FPUR1,FPUR2,ReuveniPRL,FPUR4,FPUR5,PalReuveniPRL,FPUR7,FPUR8,FPUR9,FPUR10}, e.g., foraging animals and birds come back to their dens and nests repeatedly \cite{forage,HomeRangeSearch1}; and inconvenient weather may force a team of rescuers to temporarily stop their search efforts and return to base \cite{PalReuveniPRL}. Natural disasters and epidemics may also lead to resetting, e.g., by drastically reducing the population of a living species in a certain locality \cite{population1}; and stock market crashes have a similar effect on asset prices \cite{economics}. At the microscopic level, resetting is an integral part of the well-known Michaelis-Menten reaction scheme of enzymatic catalysis \cite{Restart-Biophysics1, Restart-Biophysics2, Restart-Biophysics3, Restart-Biophysics4}. This scheme is used to describe a variety of cellular processes \cite{Gunawardena} ranging from RNA transcription \cite{Rol} to facilitated diffusion \cite{FD1,FD2} and from the work of GTPase proteins \cite{RhoA} to chaperone assisted protein folding  \cite{proteinfolding}. For all these reasons and others, resetting and its applications have now become a focal point of scientific interest \cite{ReviewSNM}. \\
\indent 
Diffusion with stochastic resetting serves as a paradigmatic model to understand resetting phenomena \cite{D1,D2,D3,D4,D5,D7,D8,D9,D10,D11,D12,D13,D14}. In this model, one considers a Brownian particle that returns to its initial position randomly in time. This resetting process drives the system away from equilibrium, giving rise to a nonequilibrium steady state and interesting relaxation dynamics. In the presence of an absorbing boundary, this model also provides a classical example of a system where resetting accelerates the completion of a first-passage process \cite{FPT1,FPT2}. \\
\indent
In many cases, diffusion occurs in the presence of a bias. A natural way to model such phenomena is to consider a Brownian particle in a suitable potential landscape. Similar to free-diffusion, resetting then leads to a nonequilibrium steady state \cite{potential1}. The scenario gets more interesting in the presence of an absorbing boundary since resetting then plays a dual role: either facilitating or hindering the resulting first-passage process \cite{PalReuveniPRL}. Moreover, as system parameters are varied, resetting can invert its role thereby leading to a resetting transition \cite{Restart-Biophysics1,Restart-Biophysics2,FPUR1}. This transition, as well as the emergence of non-equilibrium steady states, were recently explored for diffusion in various potential landscapes, e.g., linear, harmonic and power-law \cite{potential1,Landau,RayReuveniJPhysA,exponent,potential2,potential3}. An important potential landscape that was not considered despite its centrality is the logarithmic potential. \\
\indent
The logarithmic potential arises as an effective potential in a large variety of problems in Chemical, Statistical and Biological Physics. For example, in the denaturation process of double-stranded DNA, it appears as an entropic term in the free energy cost of unzipping DNA base-pairs that generate denaturation bubbles \cite{denaturation1,denaturation2,denaturation3,denaturation4,denaturation5,denaturation6}. Diffusion in an effective logarithmic potential is a popular model to study the spreading of momenta of cold atoms trapped in optical lattices \cite{opticallattice1,opticallattice2,opticallattice3,opticallattice4,opticallattice5,opticallattice6,opticallattice7,opticallattice8}. The interactions of colloids and polymers with walls of narrow channels and pores give rise to an ``entropic'' potential that is logarithmic \cite{ entropicpotential1,entropicpotential2,entropicpotential3,entropicpotential4,entropicpotential5,entropicpotential6,entropicpotential7}. Log potential also arises in Dyson's Brownian motion, where Coulomb gas is interpreted as a dynamical system to explore the eigenvalues of random matrices \cite{dysonBM1,dysonBM2}. Other well-known examples include systems with long-range interacting particles \cite{longrange}, self-gravitating Brownian particles \cite{selfgravitating1,selfgravitating2}, charges in the vicinity of a polyelectrolyte \cite{polyelectrolyte1,polyelectrolyte2} and interacting tracers in one-dimensional driven lattice gases \cite{tracerdynamics1}. \\
\indent
The logarithmic potential owns a central singularity at the origin, but behaves as a slowly varying function far away from it. Because of these unique features, diffusion in logarithmic potential leads to slow relaxation \cite{opticallattice3,opticallattice4,log1,log2} and non-trivial first-passage properties \cite{AJBray,besselFPT,ekaterina}. A Brownian particle in a logarithmic potential is therefore expected to serve as an interesting model to study the effects of resetting on stochastic dynamics. \\
\indent
In strongly repulsive logarithmic potential, a diffusing particle cannot reach the origin\cite{AJBray}. Stochastic resetting is then expected to lead to a non-equilibrium steady state. In stark contrast, when the logarithmic potential is either attractive or weakly repulsive, a diffusing particle will reach the origin in finite time \cite{AJBray}. Stochastic resetting can then either accelerate or delay the resulting first passage process. Going from the strongly repulsive case to the attractive case can be done by tuning a single model parameter --- the strength of the potential in units of the thermal energy. Diffusion with resetting in a logarithmic potential thus lends itself as attractive model to study stochastic resetting and the range response to it. In what follows, we provide a detailed analysis of this model system.\\
\indent
The rest of this paper is organized as follows. We start in Sec. II where we revisit some earlier works and review the problem of diffusion in a logarithmic potential. In Sec. III, we explore the properties of the nonequilibrium steady state that the particle attains due to resetting while it diffuses in a strongly repulsive logarithmic potential. Considering the potential to be attractive/weakly repulsive, in Sec. IV we study the first-passage of the particle in the presence of resetting to an absorbing boundary placed at the origin. In the same section, we also explore the resetting transition. The final conclusions are drawn in Sec. V, where we construct a full phase diagram for the present problem.
\section{Diffusion in a logarithmic potential}
\indent
Diffusion in a logarithmic potential has attracted considerable attention in recent years \cite{opticallattice1,opticallattice2,opticallattice3,opticallattice4,opticallattice5,opticallattice6,log1,log2,AJBray,besselFPT,ekaterina}. Here, we present a brief review of the problem. Consider a particle undergoing diffusion in a potential $U(x)=U_0\log|x|$, where $U_0$ is a constant with dimensions of energy and $x$ is dimensionless. The potential is attractive for $U_0>0$ while it is repulsive for $U_0<0$. The diffusion constant $D$ is given by the Einstein-Smoluchowski relation $D=(\beta \zeta)^{-1}$, where $\beta$ is the thermodynamic $beta$ and $\zeta$ stands for the friction coefficient. An absorbing boundary is placed at the origin such that the particle, starting from a position $x_0>0$, diffuses in the interval $0\le x<\infty$ until it hits the origin and is immediately removed from the system.\\
\indent
The time evolution of $p(x,t)$, the conditional probability density of finding the particle at position $x$ at time $t$ provided that the initial position was $x_0$, is then given by the Fokker-Planck equation
\begin{eqnarray}
\dfrac{\partial p(x,t)}{\partial t}=\dfrac{\partial }{\partial x}\left[\left(\frac{U_0}{\zeta x}\right)p(x,t)\right]
+D\dfrac{\partial^2 p(x,t)}{\partial x^{2}},
\label{eq:fpe}
\end{eqnarray}
where the initial condition is $p(x,0)=\delta(x-x_0)$ and the boundary condition reads $p(0,t)=0$. \eref{eq:fpe} is exactly solvable \cite{AJBray}. To present the solution, first we define the parameter
\begin{align}
\nu\coloneqq\frac{1+\beta U_0}{2}.
\label{eq:nu_def}
\end{align}
The solution to \eref{eq:fpe} then reads \cite{AJBray,ekaterina},
\begin{align}
p(x,t)=
\begin{cases}
\frac{x}{2Dt}\left(\frac{x_0}{x}\right)^{\nu}\exp{\left(-\frac{x^2+x_0^2}{4Dt}\right)}I_{-\nu}\left(\frac{xx_0}{2Dt}\right)\;\;\mbox{if}\;\; \beta U_0 <-1,\\  \\
\frac{x}{2Dt}\left(\frac{x_0}{x}\right)^{\nu}\exp{\left(-\frac{x^2+x_0^2}{4Dt}\right)}I_{\nu}\left(\frac{xx_0}{2Dt}\right)\;\;\;\mbox{if}\;\; \beta U_0 \ge-1,
\end{cases}
\label{eq:fpe_solution}
\end{align}
\noindent
where $I_{\pm\nu}\left(\cdot\right)$ is the modified Bessel function of the first kind with order $\pm\nu$. \\
\indent
For $\beta U_0 <-1$, the potential is strongly repulsive. One then finds from Eq. (\ref{eq:fpe_solution}) that $p(x,t)$ is always normalized, i.e., $\int_0^{\infty}p(x,t) dx=1$. Therefore, in this case, the particle never reaches the absorbing boundary at the origin. In stark contrast, for $\beta U_0 >-1$, where the potential is either weakly repulsive or attractive, the particle eventually hits the absorbing boundary for every single realization. The survival probability $Q(t)\coloneqq\int_0^{\infty}p(x,t)dx$, i.e., the probability that the particle is not absorbed at the origin by time $t$, is then given by \cite{ekaterina}
\begin{align}
Q(t)=1-\frac{\Gamma{\left(\nu,\frac{x_0^2}{4Dt}\right)}}{\Gamma(\nu)},
\label{eq:survival}
\end{align}
\noindent
where $\Gamma(\nu)\coloneqq \int_{0}^{\infty}x^{\nu-1}e^{-x}dx$ and $\Gamma(\nu,a)\coloneqq \int_{a}^{\infty}x^{\nu-1}e^{-x}dx$ denote the Gamma function and the upper incomplete Gamma function, respectively. The probability density function for the first-passage time, $T$, to the origin can be calculated from the survival probability as $f_T(t)= -\partial Q(t)/\partial t$, which gives \cite{AJBray,ekaterina} 
\begin{align}
f_T(t)=\frac{t^{-(\nu+1)}}{\Gamma{(\nu)}} \left(\frac{x_0^2}{4D}\right)^{\nu}\exp{\left(-\frac{x_0^2}{4Dt}\right)}.
\label{eq:fpt_distribution}
\end{align}
Note that in the long time limit $f_T(t)\propto t^{-(\nu+1)}$. Consequently in the same limit $Q(t)\propto t^{-\nu}$. Thus $\nu$ is known as the ``persistence exponent"\cite{FPT2}, governing the decay of the survival probability. For free diffusion $U_0=0$, and  \eref{eq:nu_def} gives $\nu=1/2$. \eref{eq:fpt_distribution} then reduces to the well-known form $f_T(t)=\frac{x_0}{\sqrt{4\pi Dt^3}} \exp{\left[-\frac{x_0^2}{4Dt}\right]}$, which describes the first-passage time (FPT) distribution of a freely diffusing particle to the absorbing boundary \cite{Cox-Miller-Book}. \\
\begin{figure}[t!]
\begin{centering}
\includegraphics[width=8.3cm]{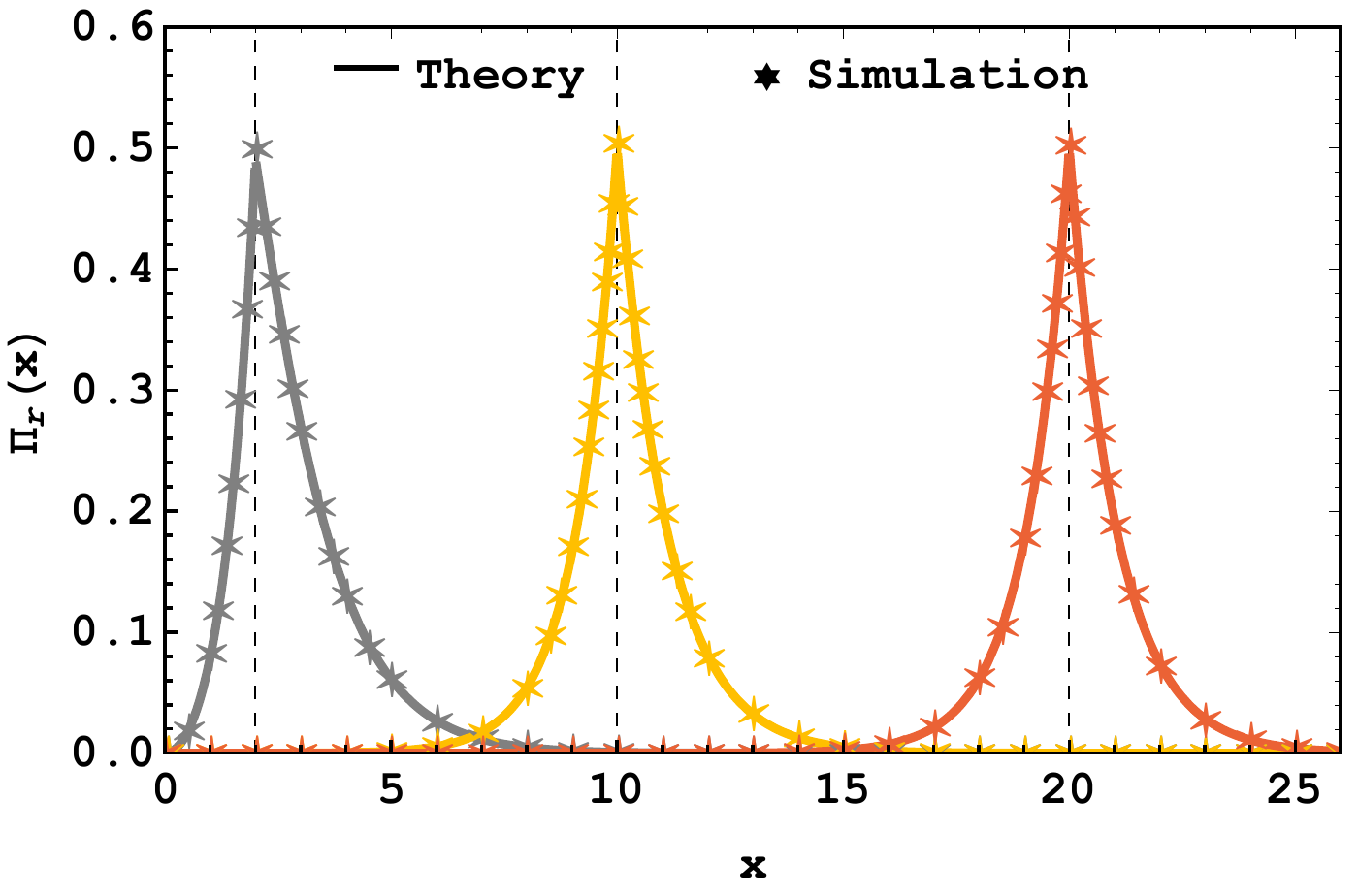}
\end{centering}
\caption{The steady state distribution for diffusion in a strongly repulsive logarithmic potential under stochastic resetting. The theoretically predicted density $\Pi_{r}(x)$ from \eref{eq:steady_state} is plotted for different values of the initial/resetting position $x_0=\{2,10,20\}$. The symbols showing simulation data are in excellent agreement with the theory. Here we have taken $\beta U_0=-2.0$ and $\alpha_0=1.0$.}
\label{Fig1}
\end{figure}
\indent
The fact that the system behaves differently depending on the potential makes the current problem a highly interesting one to study the effect of stochastic resetting. On one hand, when the particle diffuses in a strongly repulsive logarithmic potential ($\beta U_0<-1$), its survival probability in the interval $[0,\infty]$ does not decay with time. Introduction of stochastic resetting is then expected to lead to a non-equilibrium steady state. On the other hand, when the particle diffuses in either attractive or weakly repulsive logarithmic potential ($\beta U_0>-1$), its survival probability decays with time. Starting from $x_0>0$, the particle will now reach the origin in finite time. Introduction of resetting can then either expedite or delay completion of this first-passage process\cite{PalReuveniPRL}. Motivated by these different possible outcomes, we study the effects of stochastic resetting on diffusion in a logarithmic potential.
\section{Diffusion with resetting in strongly repulsive logarithmic potential}
In this Section, we explore the effect of stochastic resetting on a particle diffusing in a strongly repulsive logarithmic potential ($\beta U_0 <-1$). We assume that the particle is reset, i.e., taken back to its initial position $x_0$, with a constant rate $r$. This means that the times between two consecutive resetting events are taken from an exponential distribution with mean $1/r$. Setting $p_r(x,t)$ as the conditional probability density of finding the particle at position $x$ at time $t$ provided that the initial position was $x_0$, the Fokker-Planck equation for the process with resetting reads
\begin{eqnarray}
\dfrac{\partial p_r(x,t)}{\partial t}=
\dfrac{\partial }{\partial x}\left[\left(\frac{U_0}{\zeta x}\right) p_r(x,t)\right]
+D\dfrac{\partial^2 p_r(x,t)}{\partial x^{2}}\nonumber \\
-r p_r(x,t)+r\delta(x-x_0),
\label{eq:fper}
\end{eqnarray}
where $\delta(x-x_0)$ is a Dirac delta function. Note that in the absence of resetting ($r=0$), \eref{eq:fper} boils down to \eref{eq:fpe}, the Fokker-Planck equation for the underlying process. For $r>0$, there is a loss of probability from position $x$ and a gain of probability at position $x_0$ due to resetting. The last two terms on the right hand side of \eref{eq:fper} account for this additional probability flow, which is proportional to $r$, the resetting rate. 
\subsection{Non-equilibrium steady state}
In what follows, we will be primarily interested in the steady state probability density of the process under stochastic resetting, $\Pi_{r}(x)\coloneqq\lim_{t\rightarrow\infty}p_r(x,t)$. Since $d \Pi_r(x)/d t =0$, \eref{eq:fper} gives
\begin{eqnarray}
\dfrac{d^2 \Pi_{r}(x)}{d x^{2}}+
\dfrac{d }{d x}\left[\left(\frac{\beta U_0}{ x}\right) \Pi_{r}(x)\right]
-\left(\frac{r}{D}\right)\Pi_{r}(x)= \nonumber\\
-\left(\frac{r}{D}\right)\delta(x-x_0).
\label{eq:fper_ss}
\end{eqnarray}
The conventional way to calculate $\Pi_r(x)$ is to solve \eref{eq:fper_ss}. We discuss this in detail in Appendix A. However, the solution can also be obtained directly by using a general relation that connects the propagator of an underlying stochastic process $p(x,t)$ with its steady state distribution under resetting, $\Pi_{r}(x)$. This relation is given by \cite{ReviewSNM}
\begin{align}
\Pi_{r}(x)=\int_0^{\infty}re^{-rt}p(x,t)dt=r\tilde{p}(x,r),
\label{eq:renewal}
\end{align}
where $\tilde{p}(x,s):=\int_0^{\infty}e^{-st}p(x,t)dt$ denotes the Laplace transform of $p(x,t)$.\\
\begin{figure}[t!]
\begin{centering}
\includegraphics[width=8.5cm]{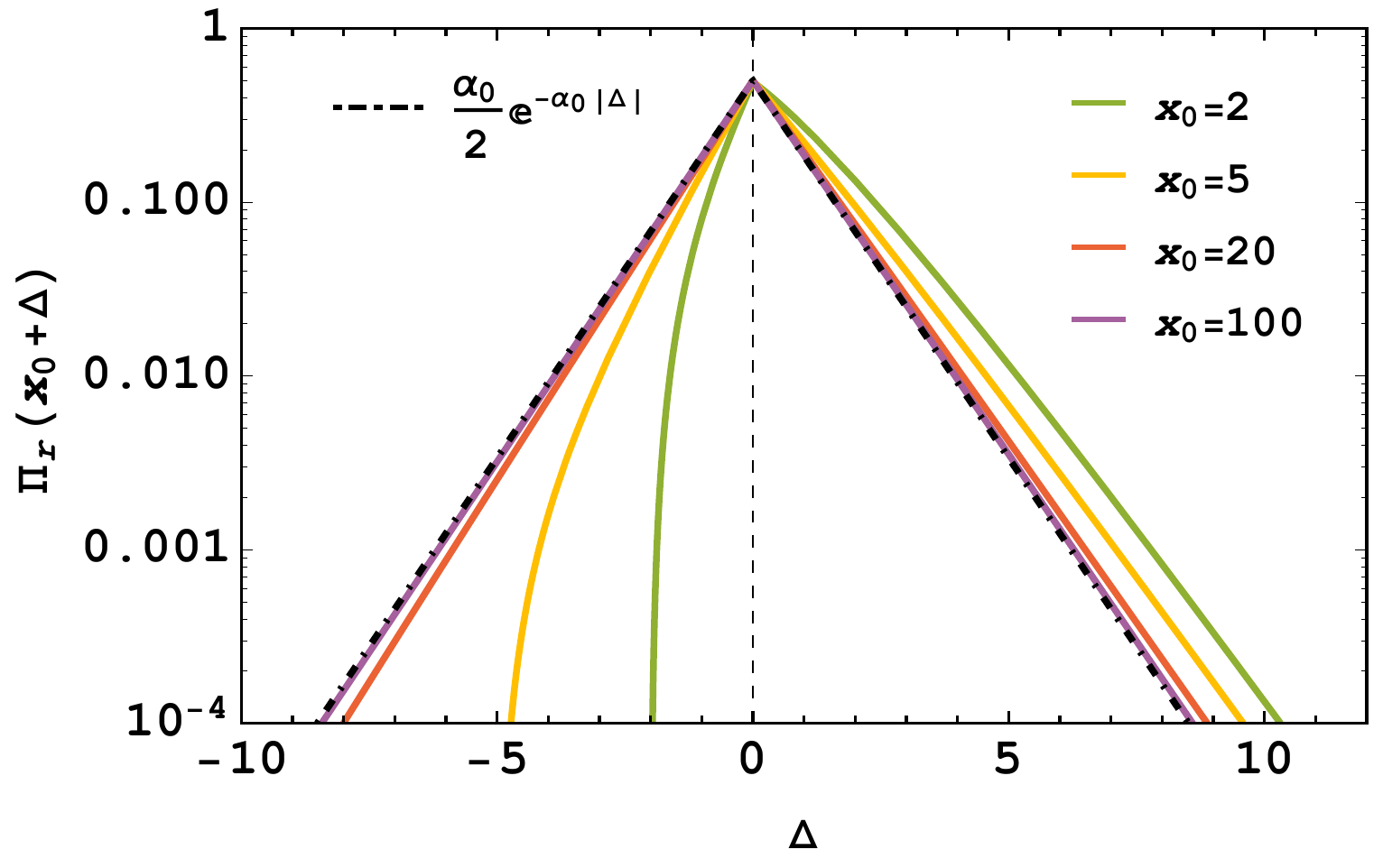}
\end{centering}
\caption{The shifted steady state distribution $\Pi_{r}(x_0+\Delta)$ for different values of $x_0$, where $\Delta \in \{-x_0,\infty\}$. The solid lines are plotted using \eref{eq:steady_state}. The dot-dashed black line shows the asymptotic result of \eref{eq:asym_highx0}. Here we have taken $\beta U_0=-4.0$ and $\alpha_0 = 1.0$.}
\label{Fig2}
\end{figure}
\indent
Therefore, in order to calculate $\Pi_{r}(x)$, we Laplace transform \eref{eq:fpe_solution} for $\beta U_0<-1$ [Appendix B] and set $s=r$ to obtain 
\begin{align}
\tilde{p}(x,r)=\begin{cases}
\frac{x}{D}\left(\frac{x_0}{x}\right)^{\nu}I_{-\nu}\left(\sqrt{\frac{rx_0^2}{D}}\right)K_{-\nu}\left(\sqrt{\frac{rx^2}{D}}\right)\;\;\mbox{if}\;\; x\geq x_0,\\ \\
\frac{x}{D}\left(\frac{x_0}{x}\right)^{\nu}K_{-\nu}\left(\sqrt{\frac{rx_0^2}{D}}\right)I_{-\nu}\left(\sqrt{\frac{rx^2}{D}}\right)\;\;\;\mbox{if}\;\; x<x_0.
\end{cases}
\label{eq:fpt_laplace}
\end{align}
Plugging \eref{eq:fpt_laplace} into \eref{eq:renewal} and setting $\alpha_0\coloneqq\sqrt{r/D}$, we get the steady state density 
\begin{align}
\Pi_{r}(x)=\begin{cases}
\alpha_0^2x\left(\frac{x_0}{x}\right)^{\nu}I_{-\nu}\left(\alpha_0x_0\right)K_{-\nu}\left(\alpha_0x\right)\;\;\mbox{if}\;x\geq x_0,\\ \\
\alpha_0^2x\left(\frac{x_0}{x}\right)^{\nu}K_{-\nu}\left(\alpha_0x_0\right)I_{-\nu}\left(\alpha_0x\right)\;\;\;\mbox{if}\;x<x_0.
\end{cases}
\label{eq:steady_state}
\end{align}
\noindent
In \fref{Fig1}, we plot $\Pi_{r}(x)$ for different values of the initial position $x_0$. The solid lines denote the analytical results, obtained by plotting \eref{eq:steady_state} and the symbols are coming from Langevin dynamics simulations. The details of the numerical simulations are given in Appendix C. \\
\indent
Examining \fref{Fig1}, we see that $\Pi_{r}(x)$ is asymmetric, which is particularly apparent for small values of $x_0$. The potential, being most repulsive at $x=0$, pushes the particle away from the origin thereby generating this asymmetry. The effect dies down for higher values of $x_0$, resulting in a steady state density that is almost symmetric.
Indeed, when $x_0\gg \alpha_0^{-1}$, the particle is most likely to stay away from the origin, where the logarithmic potential varies slowly and is almost flat. This situation is similar to free diffusion. Therefore, in this limit $\Pi_r(x)$ is expected to closely resemble a Laplace distribution, which describes the steady state density for free diffusion with stochastic resetting \cite{D1}.\\
\begin{figure}[t!]
\begin{centering}
\includegraphics[width=8.2cm]{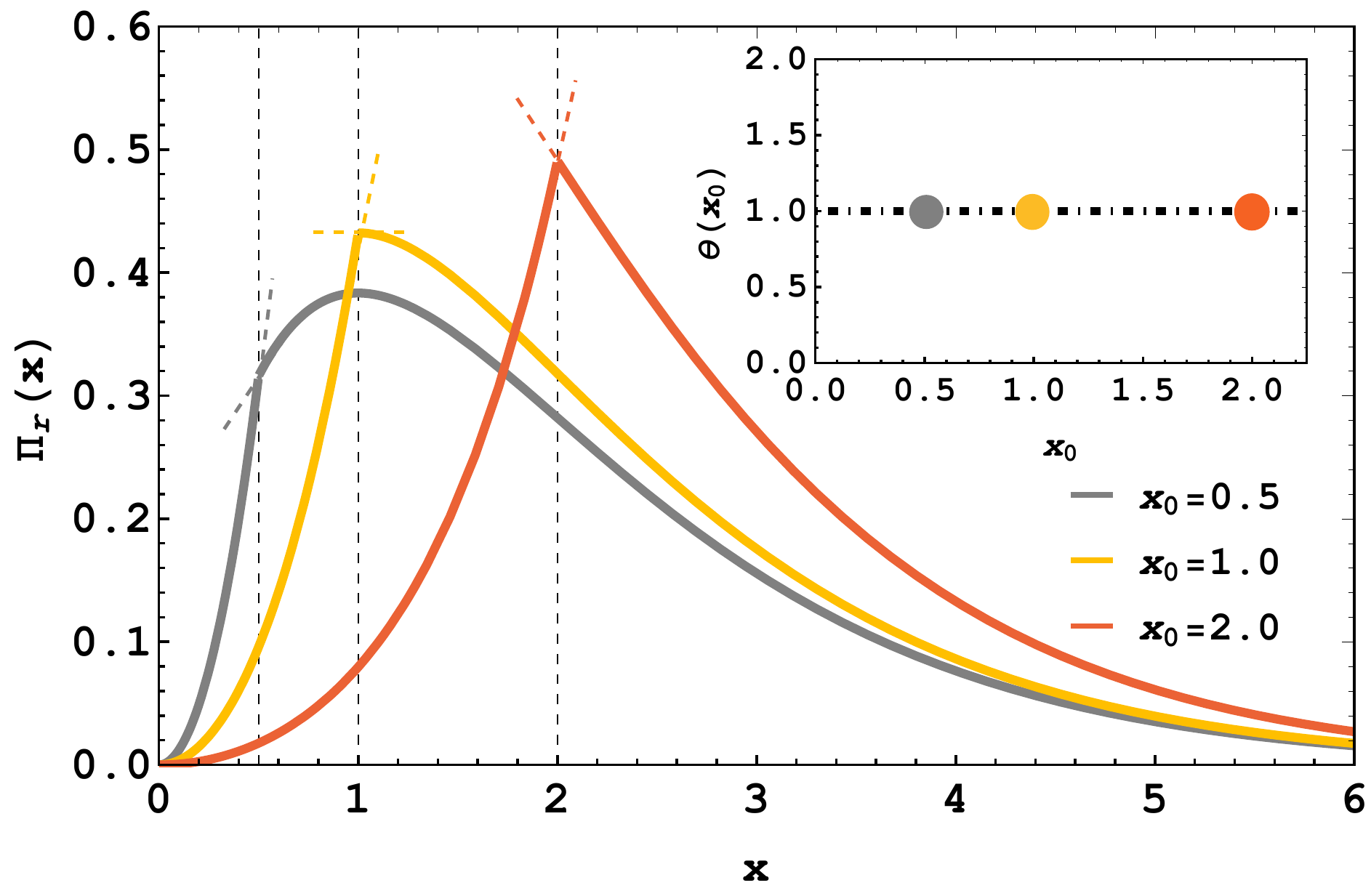}
\end{centering}
\caption{Main: The steady state density $\Pi_{r}(x)$ for different values of the initial position $x_0$. $\Pi_{r}(x)$ has a cusp at $x_0$ that coincides with its mode for moderate to high values of $x_0$. For small values of $x_0$, the mode of the distribution appears to the right of the cusp. Inset: The discontinuity in $\partial \Pi_{r}(x)/\partial x$ at $x=x_0$, denoted $\theta(x_0)$ [\eref{eq:slope_difference}], is a non-zero constant, $\alpha_0^2$. Here we have taken $\beta U_0=-2.0$ and $\alpha_0=1$.}
\label{Fig3}
\end{figure}
\indent
In order to explore this further, we set $x=x_0+\Delta$ such that $|\Delta|/x_0\ll 1$, and perform an asymptotic analysis of \eref{eq:steady_state} in the limit $x_0\gg \alpha_0^{-1}$. 
The limiting expressions of the modified Bessel functions for large arguments are $\lim_{y\rightarrow\infty}I_{-\nu}(y)\simeq e^y/\sqrt{2\pi y}$ and $\lim_{y\rightarrow\infty}K_{-\nu}(y)\simeq e^{-y}\sqrt{\pi/2y}$, respectively \cite{NIST}. Utilizing these together with \eref{eq:steady_state} we get 
\begin{eqnarray}
   \Pi_{r}(x)&=& \Pi_{r}(x_0+\Delta) \nonumber\\
   &\mbox{\stackunder[2pt]{$\simeq$}{$\scriptscriptstyle x_0\gg\alpha_0^{-1}$}}&
     \frac{\alpha_0}{2}e^{-\alpha_0 |\Delta|}\left(\frac{x_0}{x_0+\Delta}\right)^{\nu-1}\frac{x_0}{\sqrt{ x_0\left(x_0+\Delta\right)}},  \nonumber \\
   &\simeq&\frac{\alpha_0}{2}e^{-\alpha_0 |x-x_0|},
\label{eq:asym_highx0}
\end{eqnarray}
where we have neglected all terms of order $|\Delta|/x_0\ll 1$. \eref{eq:asym_highx0} proves that when $x_0\gg \alpha_0^{-1}$, the steady state distribution near $x_0$ converges to the Laplace distribution. In \fref{Fig2}, we plot the shifted steady state density $\Pi_{r}(x_0+\Delta)$ vs. $\Delta$ to explicitly show that it merges with that of free diffusion \cite{D1} for sufficiently high values of $x_0$.
\subsection{The cusp and the mode}
It is apparent from \fref{Fig1} and \ref{Fig2} that the steady state density has a cusp at the resetting position $x_0$. In what follows, we explore the cusp and the mode of the distribution in detail. In \fref{Fig3}, we plot $\Pi_{r}(x)$ for some moderate values of $x_0$, highlighting the slopes from both sides of the cusp. Letting $\Pi_{r}^-(x)$ and $\Pi_{r}^+(x)$ denote the left and right branches of $\Pi_{r}(x)$, we obtain explicit expressions for these slopes
\begin{eqnarray}
\left.  \frac{\partial \Pi_{r}^+(x)}{\partial x}  \right|_{x\to x_0}&=&\alpha_0^2I_{-\nu}(\alpha_0x_0)\left[K_{-\nu}(\alpha_0x_0)-\alpha_0x_0K_{-1-\nu}(\alpha_0x_0)\right],\nonumber\\
\left. \frac{\partial \Pi_{r}^-(x)}{\partial x} \right|_{x\to x_0}&=&\alpha_0^2K_{-\nu}(\alpha_0x_0)\left[I_{-\nu}(\alpha_0x_0)+\alpha_0x_0I_{-1-\nu}(\alpha_0x_0)\right].\nonumber\\
\label{eq:slope_pdf}
\end{eqnarray}
\noindent
The difference between the slopes at $x_0$ in \eref{eq:slope_pdf} then reads
\begin{eqnarray}
\theta(x_0)=\left[\frac{\partial \Pi_{r}^-(x)}{\partial x} -\frac{\partial \Pi_{r}^+(x)}{\partial x}\right]_{x\to x_0}=\alpha_0^2,
\label{eq:slope_difference}
\end{eqnarray}
where we utilized the identity\cite{NIST} $K_{-\nu}(y)I_{-1-\nu}(y)+I_{-\nu}(y)K_{-1-\nu}(y)=1/y$. Thus, regardless of the particle's initial position, the difference between the slopes at $x_0$ is always $\alpha_0^2$, as is illustrated in the inset of \fref{Fig3}. This non-zero difference proves that $\Pi_r(x)$ always has a cusp at $x_0$. Note that \eref{eq:slope_difference} follows directly from \eref{eq:fper_ss}, as we show in Appendix A. \\
\begin{figure}[t!]
\begin{centering}
\includegraphics[width=8.4cm]{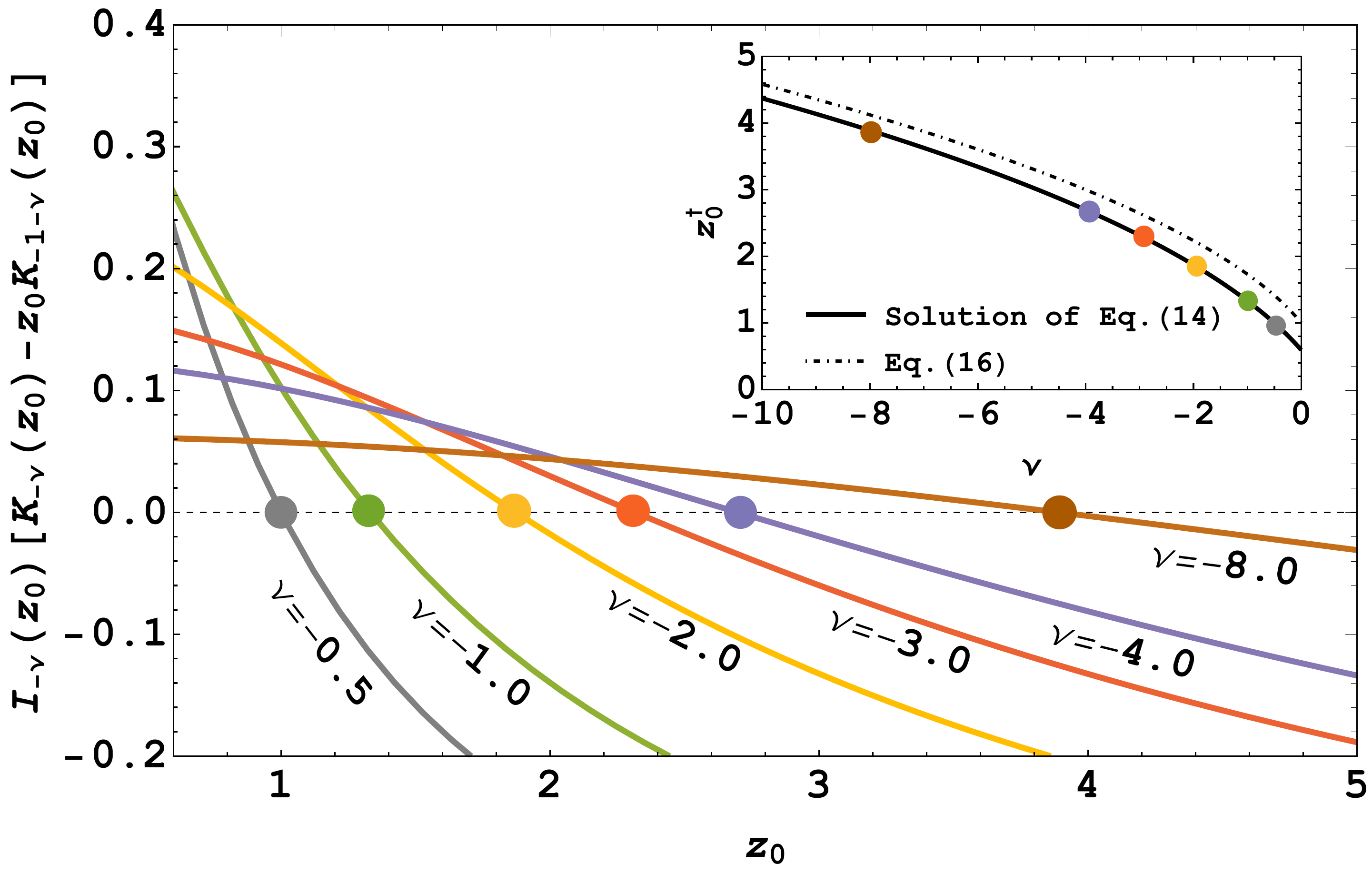}
\end{centering}
\caption{Main: Graphical solution of \eref{eq:Cusp_TranscendentalEq} for different values of $\nu$. The solutions, $z_0^{\dagger}$, are highlighted by circles. Inset: Solid line shows the variation of $z_0^{\dagger}$ vs. $\nu$, with colored circles corresponding to the values of $\nu$ from the main figure. The dashed line shows the upper bound from \eref{eq:UpperBound}.}
\label{Fig4}
\end{figure}
\indent
It is evident from \fref{Fig3} that for moderate to high values of $x_0$, the mode of the distribution coincides with the cusp, whereas for small values of $x_0$, the mode appears to the right of the cusp. This effect arises due to the interplay between resetting and the action of the logarithmic potential. On one hand, resetting takes the particle back to $x_0$ and thus increases the probability density of finding it there. On the other hand, the strongly repulsive potential pushes the particle away from the origin, thereby decreasing the probability density of finding it close to $x=0$. For moderate to high values of $x_0$, the effect of resetting dominates: the cusp and the mode of $\Pi_r(x)$ coincide. In contrast, for small values of $x_0$ repulsion plays a significant role pushing the mode of the distribution to the right of the cusp.\\
\begin{figure}[t!]
\begin{centering}
\includegraphics[width=8.0cm]{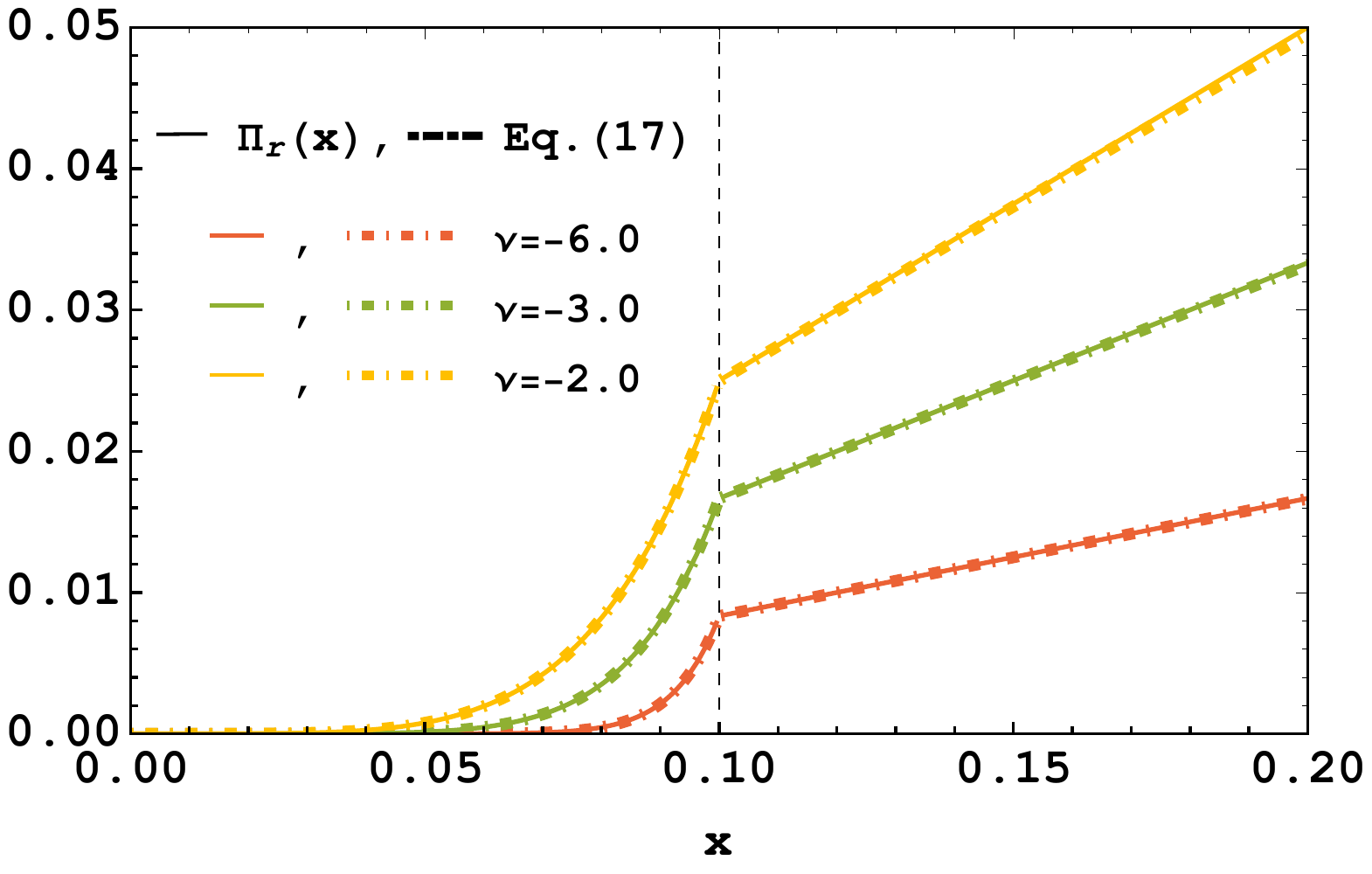}
\end{centering}
\caption{The steady state distribution for different values of $\nu$ in the limit $x_0\ll \alpha_0^{-1}$. The solid lines denote $\Pi_{r}(x)$ from \eref{eq:steady_state}. The dash-dotted lines show limiting results from \eref{eq:asym_low_x0_low_x}. Here we have taken $x_0=0.1$ and $\alpha_0 = 1.0$. }
\label{Fig5}
\end{figure}
\indent
In what follows, we will be interested in finding the minimal value of $x_0$, denoted $x_0^{\dagger}$, for which the cusp and the mode of $\Pi_r(x)$ coincide. To do that, we see from \fref{Fig3} that while the slope of $\Pi_{r}^-(x)$ is always positive at $x_0$, that of $\Pi_{r}^+(x)$ changes its sign. This transition happens at $x_0^{\dagger}$, where the slope of $\Pi_{r}^+(x)$ is exactly zero. Setting $\partial \Pi_{r}^+(x)/\partial x=0$ and utilizing \eref{eq:slope_pdf} gives the following transcendental equation
\begin{align}
I_{-\nu}(z_0)\left[K_{-\nu}(z_0)-z_0K_{-1-\nu}(z_0)\right]=0,
\label{eq:Cusp_TranscendentalEq}
\end{align}
where $z_0\coloneqq \alpha_0x_0$. The solution of \eref{eq:Cusp_TranscendentalEq}, denoted $z_0^{\dagger}$, can then be used to calculate the transition point
\begin{align}
x_0^{\dagger} = z_0^{\dagger}/\alpha_0.
\label{eq:z0_def}
\end{align}
In \fref{Fig4}, we graphically solve \eref{eq:Cusp_TranscendentalEq} for different values of $\nu$. In the inset, we plot the solutions, $z_0^{\dagger}$, vs $\nu\leq 0$. We find $z_0^{\dagger}$ to be a monotonically decreasing function of $\nu$. Thus recalling \eref{eq:nu_def}, we conclude that as the potential becomes more repulsive, the transition point $x_0^{\dagger}$ is pushed further away from the origin.\\
\indent
From \eref{eq:Cusp_TranscendentalEq}, we see that at the transition point, $K_{-\nu}(z_0^{\dagger})/K_{-1-\nu}(z_0^{\dagger})=z_0^{\dagger}$, since $I_{-\nu}(z_0^{\dagger})\neq0$ for $z_0^{\dagger}>0$. The ratio of the modified Bessel functions of the second kind satisfies the inequality \cite{BesselBound} $K_{-\nu}(y)/K_{-1-\nu}(y)<(-\nu +\sqrt{\nu^2+y^2})/y$, leading to an upper bound on $z_0^{\dagger}$
\begin{align}
z_0^{\dagger}<\sqrt{1-2\nu}.
\label{eq:UpperBound}
\end{align}
In the inset of \fref{Fig4}, we plot this upper bound to show that \eref{eq:UpperBound} provides a simple yet effective way to locate the transition point $z_0^{\dagger}$, bypassing the solution of \eref{eq:Cusp_TranscendentalEq}.
\subsection{Asymptotic analysis}
Next, we analyze the behavior of the steady state density in the limit $x_0\ll \alpha_0^{-1}$, i.e., where $\Pi_r(x)$ is highly asymmetric. To explore this, we set $x=x_0+\Delta$ such that $|\Delta|\le x_0$ and perform an asymptotic analysis of \eref{eq:steady_state} in the limit $x_0\ll \alpha_0^{-1}$. The limiting expressions of the modified Bessel functions for small arguments \cite{NIST} read $\lim_{y\rightarrow 0}I_{-\nu}(y)\simeq 2^{\nu}y^{-\nu}/\Gamma(1-\nu)$ and $\lim_{y\rightarrow 0}K_{-\nu}(y)\simeq 2^{-1-\nu}y^{\nu}\Gamma(-\nu)$, respectively. Utilizing these along with \eref{eq:steady_state} we obtain
\begin{eqnarray}
   \Pi_{r}(x)&=& \Pi_{r}(x_0+\Delta) \nonumber\\
   &\mbox{\stackunder[2pt]{$\simeq$}{$\scriptscriptstyle x_0\ll\alpha_0^{-1}$}}&
   \begin{cases} -\frac{\alpha_0^2}{2\nu} x\;\;\;\mbox{if}\;\;\Delta\geq 0, \\ \\-\frac{\alpha_0^2x_0^{2\nu}}{2\nu} x^{1-2\nu} \;\;\;\mbox{if}\;\;\Delta<0. 
   \end{cases}
\label{eq:asym_low_x0_low_x}
\end{eqnarray}
In \fref{Fig5}, we plot the steady state distribution $\Pi_{r}(x)$ in the limit $x_0\ll \alpha_0^{-1}$ for different values of $\nu$. Agreement with \eref{eq:asym_low_x0_low_x} is clearly evident.\\
\begin{figure}[t!]
\begin{centering}
\includegraphics[width=8.4cm]{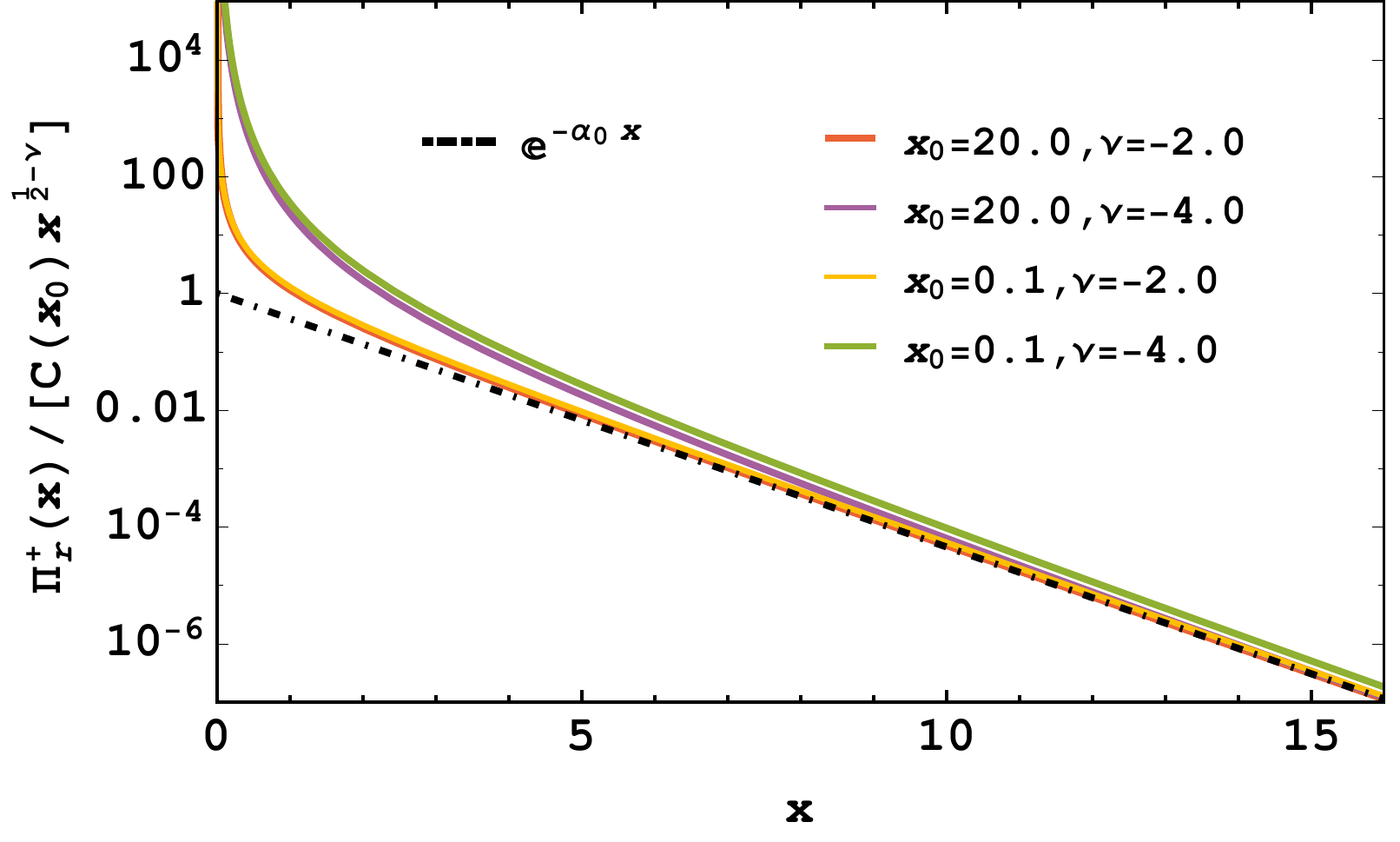}
\end{centering}
\caption{The scaled steady state distribution $\Pi_{r}^+(x)/[C(x_0)x^{\frac{1}{2}-\nu}]$ for different values of $\nu$. Here, $\Pi_{r}^+(x)$ is taken from \eref{eq:steady_state} and $C(x_0)$ is taken from \eref{eq:cx0}. The dash-dotted black line shows the limiting expression $e^{-\alpha_0 x}$ with $\alpha_0 = 1.0$. }
\label{Fig6}
\end{figure}
\indent
Finally, we explore how the steady state density decays at large values of $x$. We separate two cases; in the limit $x_0\gg \alpha_0^{-1}$, we recall that $\lim_{y\rightarrow\infty}I_{-\nu}(y)\simeq e^y/\sqrt{2\pi y}$ and  $\lim_{y\rightarrow\infty}K_{-\nu}(y)\simeq e^{-y}\sqrt{\pi/2y}$, and use \eref{eq:steady_state} to obtain
\begin{eqnarray}
   \Pi_{r}^+(x) \mbox{\stackunder[2pt]{$\simeq$}{$\scriptscriptstyle \substack{x_0\gg \alpha_0^{-1}\\ x\to \infty }$}}
   \frac{\alpha_0}{2}\left(\frac{ x}{x_0}\right)^{\frac{1}{2}-\nu}e^{-\alpha_0 (x-x_0)}.
\label{eq:symmetric_decay}
\end{eqnarray}
On the other hand, in the limit $x_0\ll \alpha_0^{-1}$, utilizing the limiting expressions\cite{NIST} $\lim_{y\rightarrow 0}I_{-\nu}(y)\simeq 2^{\nu}y^{-\nu}/\Gamma(1-\nu)$ and  $\lim_{y\rightarrow\infty}K_{-\nu}(y)\simeq e^{-y}\sqrt{\pi/2y}$ together with \eref{eq:steady_state}, we get
\begin{eqnarray}
  \Pi_{r}^+(x) \mbox{\stackunder[2pt]{$\simeq$}{$\scriptscriptstyle 
   \substack{x_0\ll \alpha_0^{-1}\\ x\to \infty }$}}
   \frac{\sqrt{\pi}\alpha_0}{\Gamma(1-\nu)}\left[\frac{\alpha_0 x}{2}\right]^{\frac{1}{2}-\nu}e^{-\alpha_0 x}.
\label{eq:asymmetric_decay}
\end{eqnarray}
Comparing \eref{eq:symmetric_decay} and (\ref{eq:asymmetric_decay}), we see that in the large $x$ limit \begin{align}
\Pi_{r}^+(x) \mbox{\stackunder[4pt]{$\simeq$}{$\scriptscriptstyle x\to\infty$}}
\;C(x_0)\;x^{\frac{1}{2}-\nu}e^{-\alpha_0 x},
    \label{eq:ssd_decay}
\end{align}
where the prefactor is given by
\begin{eqnarray}
   C(x_0)=
   \begin{cases} \frac{\alpha_0}{2} x_0^{\nu-\frac{1}{2}}e^{\alpha_0 x_0}\;\;\;\mbox{for}\;\;x_0\gg \alpha_0^{-1}, \\ \\\frac{\sqrt{\pi}\alpha_0}{\Gamma(1-\nu)}(\frac{\alpha_0}{2})^{\frac{1}{2}-\nu} \;\;\;\mbox{for}\;\;x_0\ll \alpha_0^{-1}. 
   \end{cases}
\label{eq:cx0}
\end{eqnarray}
In \fref{Fig6}, we plot the scaled steady state distribution $\Pi_{r}^+(x)/[C(x_0)x^{\frac{1}{2}-\nu}]$ to show that it always converges to the $\nu$-independent function $e^{-\alpha_0 x}$ for large values of $x$.
\section{Diffusion with resetting in weakly repulsive or attractive logarithmic potential}
In the previous section, we explored the non-equilibrium steady state attained for diffusion with resetting in a strongly repulsive logarithmic potential ($\beta U_0<-1$). In contrast, when the potential is weakly repulsive or attractive ($\beta U_0>-1$), a steady state is not achieved since the particle will eventually hit the absorbing boundary at the origin \cite{AJBray,ekaterina}. Here we explore the effect of stochastic resetting on this first passage process.
\subsection{First-passage with resetting}
Consider a particle that diffuses in a potential $U(x)=U_0\log|x|$ with $\beta U_0>-1$ and is subject to resetting with a constant rate $r$. The conventional way to calculate its first-passage to the absorbing boundary at $x=0$ involves the backward Fokker Planck equation approach\cite{D1}. In this method, one treats the initial position $x_0$ as a variable. Hence, we assume that after each resetting event, the particle returns to $x_r>0$ (which in principle can be different than its initial position $x_0>0$), solve the problem with arbitrary $x_0$ and $x_r$ and eventually set $x_r = x_0$. Recalling that $p_r(x,t|x_0)$ is the probability density of finding the particle at position $x$ at time $t$, we see that its backward Fokker Planck equation\cite{gardiner,ReviewSNM} reads
\begin{eqnarray}
\dfrac{\partial p_r(x,t|x_0)}{\partial t}&=&
-\left(\frac{U_0}{\zeta x_0}\right)\dfrac{\partial p_r(x,t|x_0)}{\partial x_0}
+D\dfrac{\partial^2 p_r(x,t|x_0)}{\partial x_0^{2}}\nonumber \\
&-&r p_r(x,t|x_0)+rp_r(x,t|x_r).
\label{eq:bme}
\end{eqnarray}
The survival probability $Q_r(t|x_0)\coloneqq\int_{0}^{\infty}p_r(x,t|x_0)dx$, i.e., the probability that the particle is not absorbed at $x=0$ by time $t$, then evolves in time following the master equation
\begin{eqnarray}
\dfrac{\partial Q_r(t|x_0)}{\partial t}=
-\left(\frac{U_0}{\zeta x_0}\right)\dfrac{\partial Q_r(t|x_0) }{\partial x_0}+D\dfrac{\partial^2 Q_r(t|x_0)}{\partial x_0^{2}}\nonumber\\-r Q_r(t|x_0)+rQ_r(t|x_r),
    \label{eq:bme_Q}
\end{eqnarray}
where the initial condition is $Q_r(0|x_0)=1$ and the boundary condition reads $Q_r(t|0)=0$. \eref{eq:bme_Q} is exactly solvable and the solution allows us to calculate the distribution of the FPT for the process under resetting, bypassing the calculation of $p_r(x,t|x_0)$. This derivation is given in Appendix D. 

An alternative way to calculate the FPT distribution is by utilizing the general theory of first-passage under resetting \cite{ReuveniPRL,PalReuveniPRL}. This theory asserts that the FPT distribution of a process with a constant resetting rate $r$ can be expressed in terms of the FPT distribution of the process without resetting. Letting $T$ denote the FPT of a generic process and $T_r$ its FPT under resetting, we set $\tilde{T}(s)\coloneqq\left<\exp{(-sT)}\right>$ and $\tilde{T}_r(s)\coloneqq\left<\exp{(-sT_r)}\right>$ as the Laplace transforms of $T$ and $T_r$, respectively. One can then show that the following relation holds \cite{ReuveniPRL}
\begin{equation}
\tilde{T}_r(s)=\frac{\tilde{T}(s+r)}{\frac{s}{s+r}+\frac{r}{s+r} \tilde{T}(s+r)}.\\
\label{eq:fptd_r}
\end{equation}
Eq.~(\ref{eq:fptd_r}) is completely general and we will now use it to analyze diffusion with resetting in weakly repulsive ($-1<\beta U_0<0$) and attractive ($\beta U_0>0$) logarithmic potential.\\
\indent
Consider first the problem without resetting. The distribution of the first passage time $T$ to the absorbing boundary at the origin is then given by \eref{eq:fpt_distribution}. Laplace transforming \eref{eq:fpt_distribution} [Appendix E] we get
\begin{equation}
\tilde{T}(s)=\frac{\left(\sqrt{s x_0^2/D}\right)^{\nu}K_{\nu}\left(\sqrt{sx_0^2/D} \right)}{2^{\nu-1}\Gamma{(\nu)}}.\\
\label{eq:fpt_lt}
\end{equation}
Here again, $K_{\nu}(\cdot)$ denotes the modified Bessel function of the second kind and $\Gamma(\nu)$ denotes the Gamma function. Plugging \eref{eq:fpt_lt} into \eref{eq:fptd_r} we obtain
\begin{equation}
\tilde{T}_r(s)=\frac{(s+r)\left(\alpha x_0\right)^{\nu}K_{\nu}\left(\alpha x_0\right)}{s\;2^{\nu-1}\Gamma{(\nu)}+ r \left(\alpha x_0\right)^{\nu}K_{\nu}\left(\alpha x_0\right)},\\
\label{eq:fpt_lt_r}
\end{equation}
where $\alpha= \sqrt{(s+r)/D}$. Finally, to get the survival, we recall that $Q_r(t|x_0)=\int_{t}^{\infty}f_{T_r}(t)dt$, with $f_{T_r}(t)$ standing for the probability density function of $T_r$. Thus, $\tilde{Q_r}(s|x_0)=[1-\tilde{T_r}(s)]/s$ which gives  
\begin{align}
\tilde{Q_r}(s|x_0)=\frac{2^{\nu-1}\Gamma(\nu)-(\alpha x_0)^{\nu}K_{\nu}(\alpha x_0)}{s\; 2^{\nu-1}\Gamma(\nu)+r\;(\alpha x_0)^{\nu}K_{\nu}(\alpha x_0)}.
    \label{eq:survival_solution_main}
\end{align}
\eref{eq:fpt_lt_r}) and \eref{eq:survival_solution_main} provide closed-form expressions for the FPT distribution and survival of our process in Laplace space.

\subsection{Mean and standard deviation}

\begin{figure*}[t!]
\begin{centering}
\includegraphics[width=4.30cm]{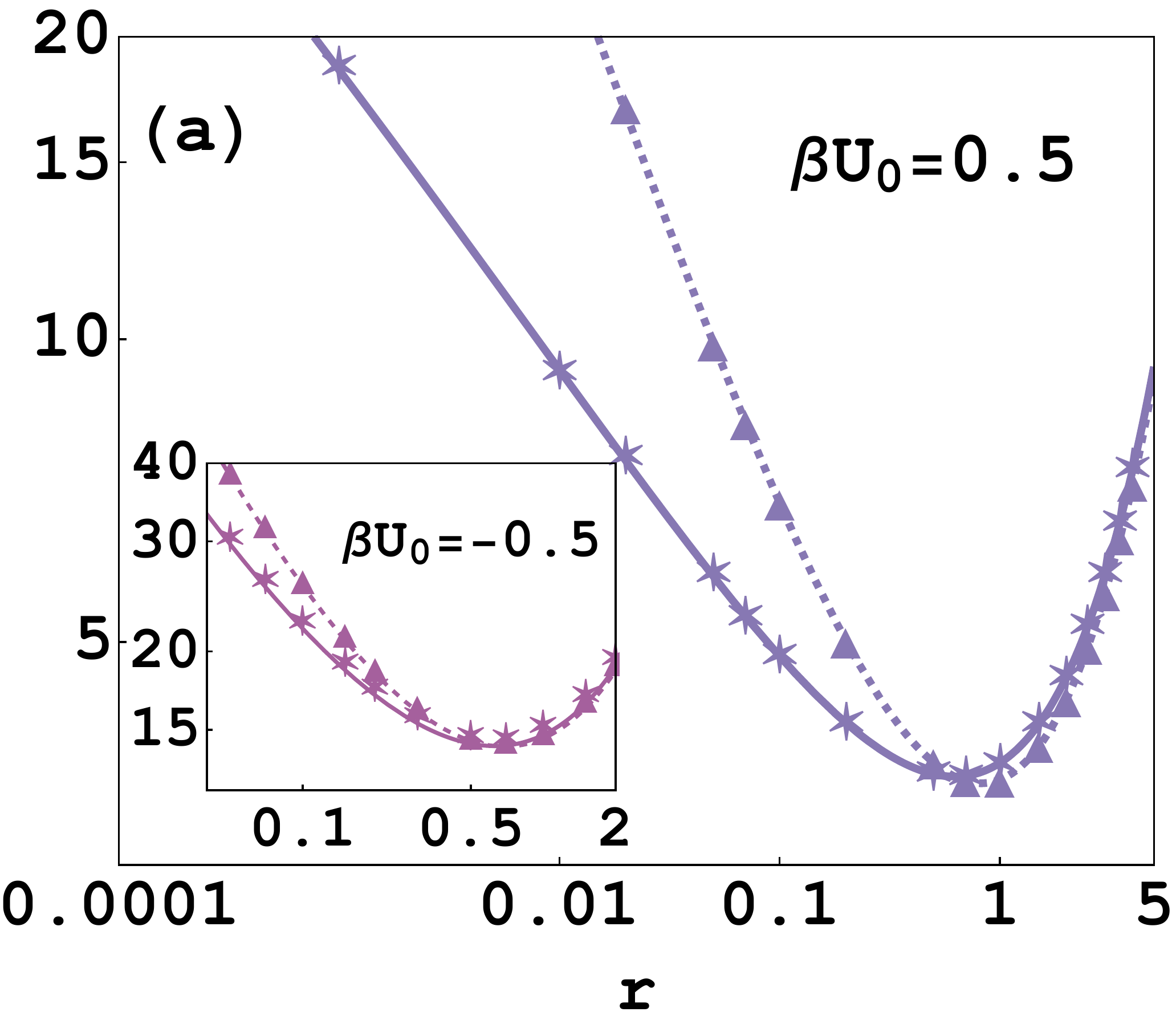}
\includegraphics[width=4.26cm]{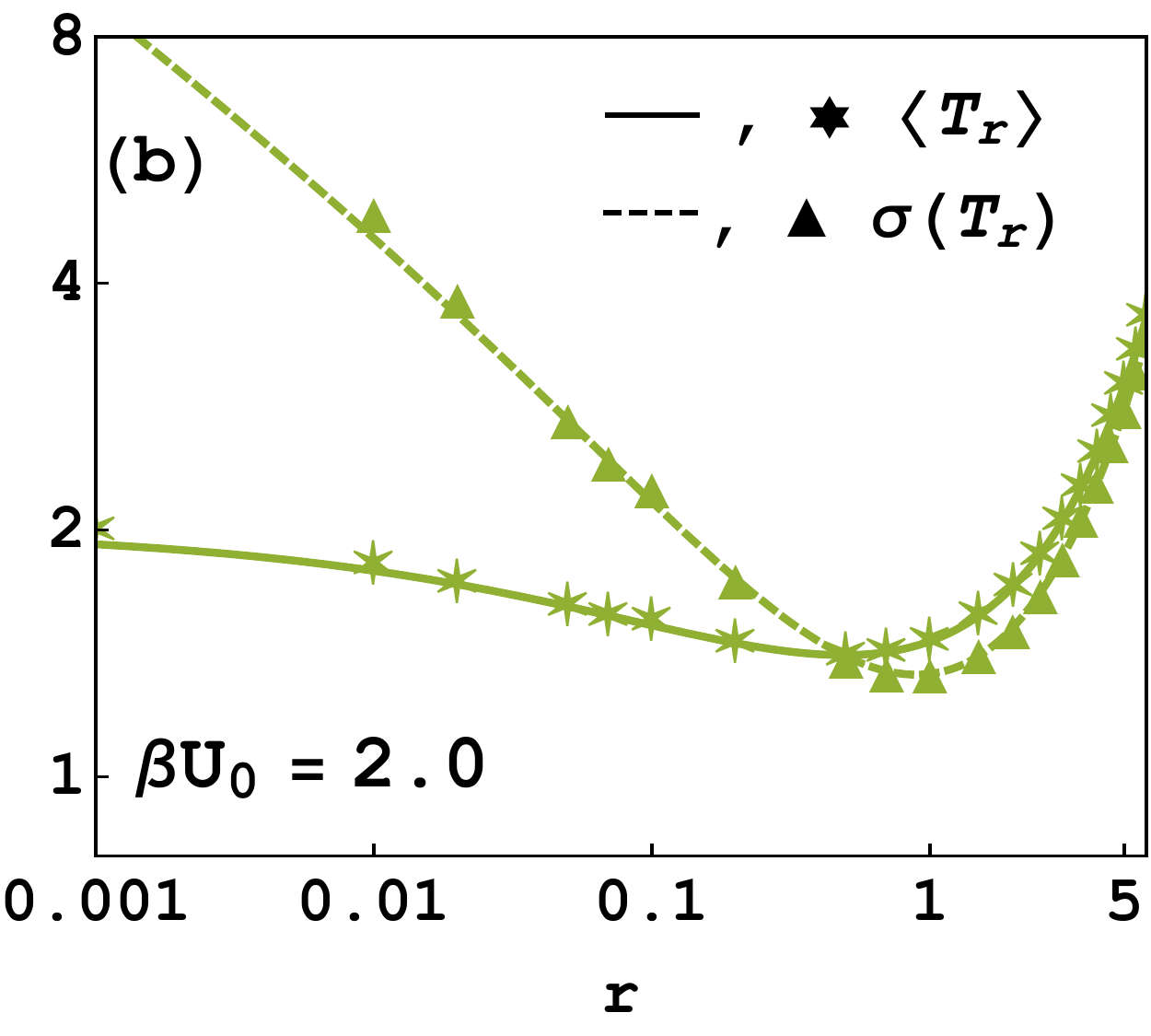}
\includegraphics[width=4.50cm]{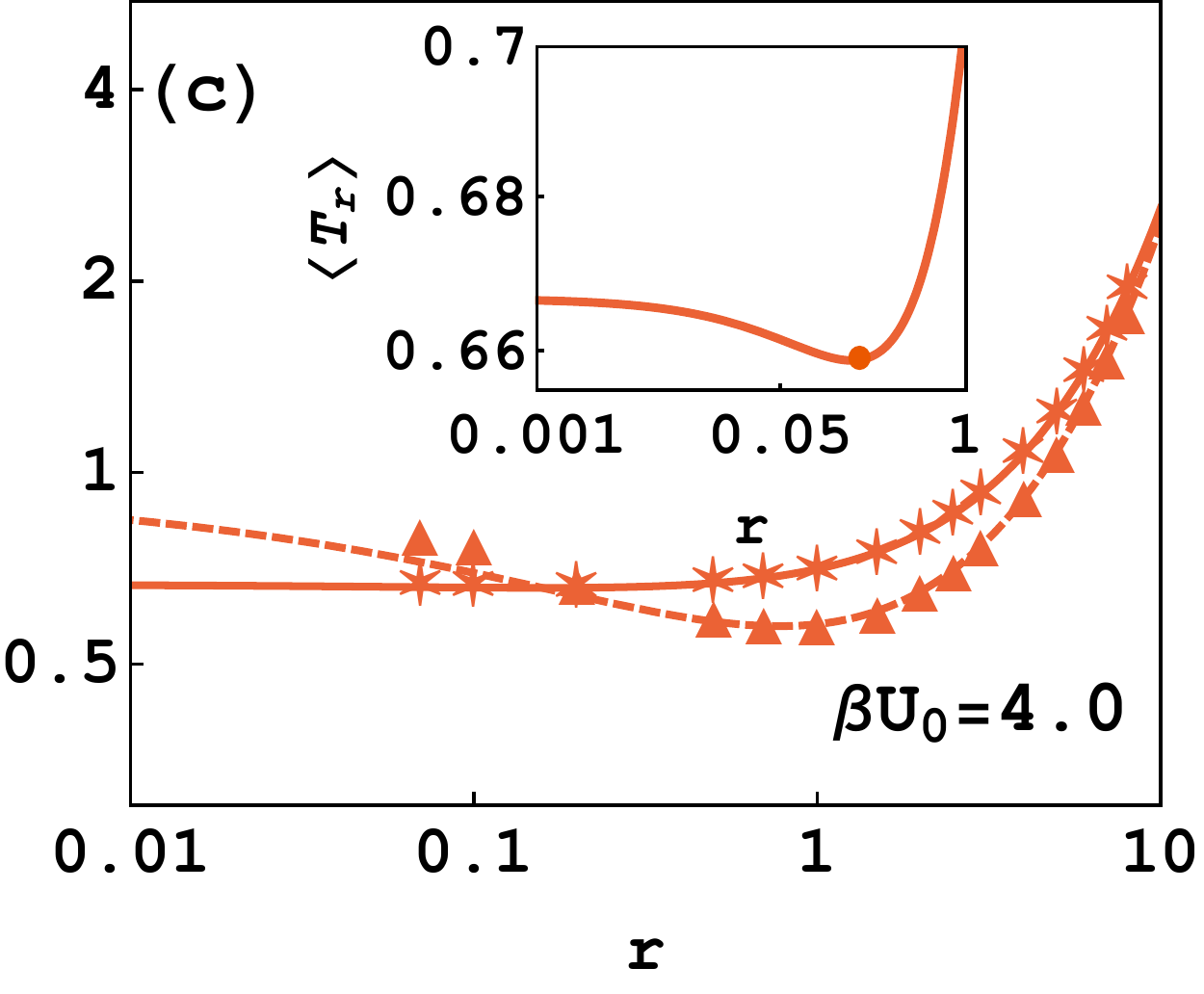}
\includegraphics[width=4.38cm]{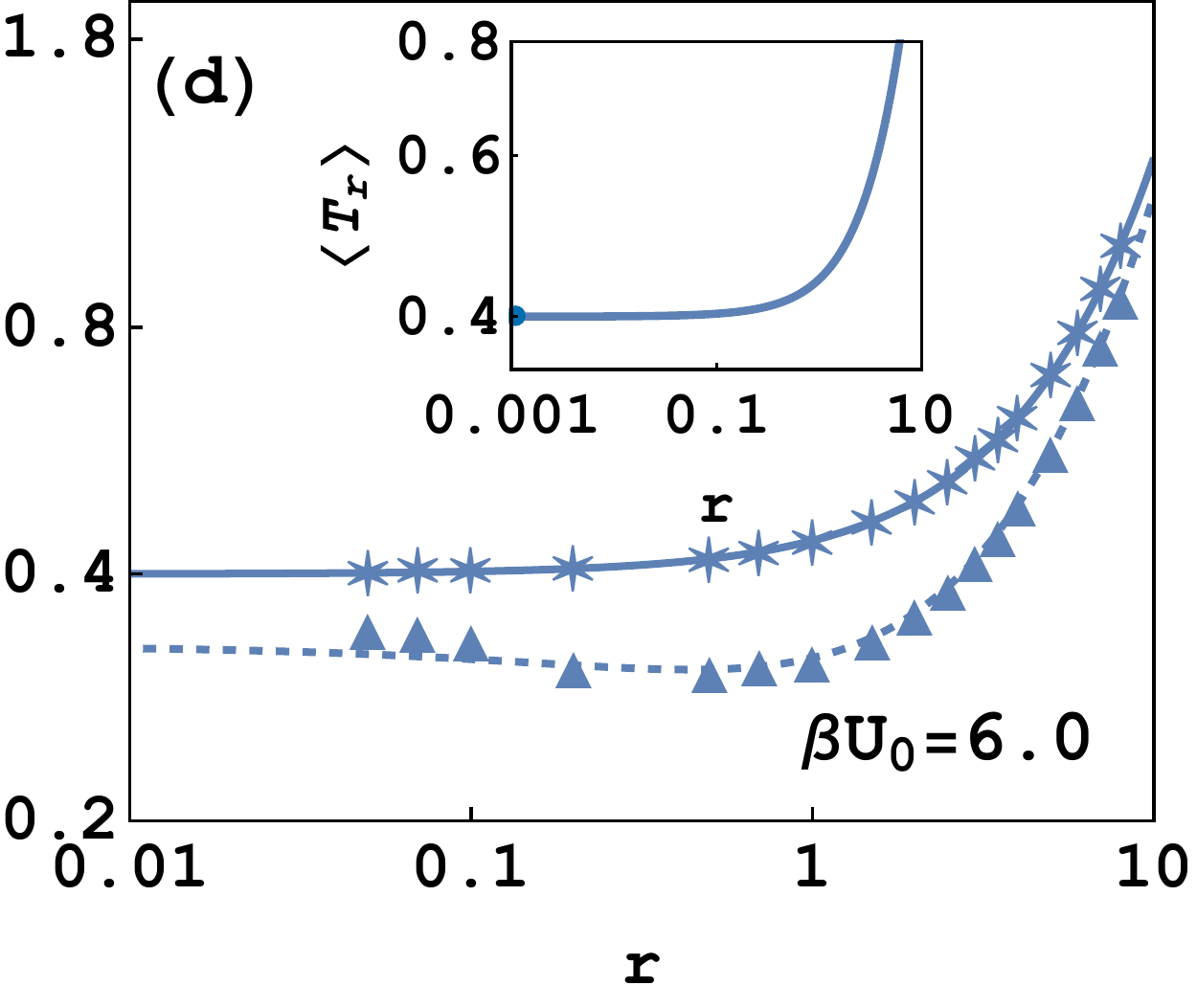}%
\end{centering}
\caption{The mean and the standard deviation of the FPT vs. the resetting rate, $r$, for four different phases of diffusion in a logarithmic potential. Lines indicate analytical results [solid lines: $\left<T_r\right>$, dashed lines $\sigma(T_r)$] and symbols show data from simulations [stars: $\left<T_r\right>$, triangles: $\sigma(T_r)$]. Panel (a): For $-1<\beta U_0<1$, the mean $\left<T_r\right>$ and the standard deviation $\sigma(T_r)$ of the FPT show a non-monotonic variation with the resetting rate, and both diverge in the limit $r\to 0$ (no resetting). Panel (b): For $1<\beta U_0<3$, the mean and standard deviation of the FPT show a non monotonic variation with the resetting rate. However, while the mean FPT is finite in the limit  $r\to 0$, the standard deviation diverges. Panel (c): For $3<\beta U_0<5$, the mean and the standard deviation of FPT still show a non monotonic variation with the resetting rate. Both are finite in the limit $r\to 0$, and  $\sigma(T_{r=0})>\left<T_{r=0}\right>$. Panel (d): For $\beta U_0>5$, the mean FPT is monotonically increasing with the resetting rate. Both the mean and standard deviation are finite in the limit $r\to 0$, and $\sigma(T_{r=0})<\left<T_{r=0}\right>$. The insets of panels (c) and (d) present a zoom in on $\left<T_r\right>$ from the corresponding main plots. In all panels we have taken $x_0 = 2.0$ and $D=1.0$.}
\label{Fig7}
\end{figure*}

In this subsection, we compute the mean and variance of the first passage time $T_r$. With these at hand, we characterize four distinct FPT phases for diffusion with resetting in weakly repulsive or attractive logarithmic potential. We start with the mean FPT, which for a generic process under stochastic resetting follows directly from \eref{eq:fptd_r}\cite{ReuveniPRL} 
\begin{equation}
\left<T_r\right>=-[d \tilde{T}_r(s)/d s]_{s=0}=\frac{1}{r}\left[\frac{1-\tilde{T}(r)}{\tilde{T}(r)}\right]. \\
\label{eq:mfpt_lt}
\end{equation}
In our case, plugging \eref{eq:fpt_lt} into \eref{eq:mfpt_lt}, we obtain the mean FPT to the origin
\begin{align}
\left<T_r\right>=\frac{1}{r}\left[\frac{2^{\nu-1}\Gamma(\nu)}{(\alpha_0 x_0)^{\nu}K_{\nu}(\alpha_0 x_0 )}-1\right],
    \label{eq:mfpt}
\end{align}
\noindent
with $\alpha_0 = \sqrt{r/D}$, as introduced in Sec. III. Note that for $U_0=0$, i.e., free diffusion, $\nu=1/2$ [\eref{eq:nu_def}]. The mean FPT then boils down to $\left<T_r\right>=(e^{\sqrt{rx_0^2/D}}-1)/r$, which agrees with the result obtained by Evans and Majumdar for free diffusion with stochastic resetting \cite{D1}.\\
\indent
The second moment of the FPT can be calculated from \eref{eq:fptd_r} in a similar manner \cite{ReuveniPRL}
\begin{align}
\left<T_r^2\right>=[d^2 \tilde{T}_r(s)/d s^2]_{s = 0}=2\frac{r\frac{d \tilde{T}(r)}{d r}-\tilde{T}(r)+1}{r^2\left[\tilde{T}(r)\right]^2}.
    \label{eq:2nd_moment}
\end{align}
\noindent
Plugging \eref{eq:fpt_lt} into \eref{eq:2nd_moment}, one can calculate $\left<T_r^2\right>$ and that in turn gives the standard deviation 
\begin{eqnarray}
&\left.\right.&\sigma(T_r)\coloneqq\sqrt{\left<T_r^2\right>-\left<T_r\right>^2}\nonumber\\
&=&
\frac{1}{r}\sqrt{\frac{2^{\nu-1}\Gamma(\nu)\left[2^{\nu-1}\Gamma(\nu)-(\alpha_0 x_0)^{1+\nu}K_{\nu-1}(\alpha_0 x_0 )\right]}{(\alpha_0 x_0)^{2\nu}[K_{\nu}(\alpha_0 x_0 )]^2}-1}.\nonumber\\
    \label{eq:stdev}
\end{eqnarray}
\indent 
In \fref{Fig7}, we plot the mean and the standard deviation of the FPT vs. the resetting rate $r$ for different values of $\beta U_0\in\{-1,\infty\}$. We observe four distinct phases shown in panels (a), (b), (c) and (d). These show noticeably different behavior in the limit $r\to 0$. For weakly repulsive ($-1<\beta<0$) or weakly attractive ($0<\beta U_0<1$) potential, both the mean and standard deviation of the FPT diverge in the absence of resetting, i.e., for $r=0$ [panel (a)]. For weak to moderately attractive potential, where $1<\beta U_0<3$, the mean FPT in the absence of resetting is finite, but the standard deviation at $r=0$ diverges [panel (b)]. When the potential is moderately attractive ($3<\beta U_0<5$), both $\left<T_r\right>$ and $\sigma(T_r)$ become finite at $r=0$, with $\sigma(T_{r=0})>\left<T_{r=0}\right>$ [panel (c): main figure]. In this phase, the mean FPT still shows a non-monotonic behavior with the resetting rate [panel (c): inset]. Finally, for strongly attractive potential ($\beta U_0>5$), we again find that both $\left<T_r\right>$ and $\sigma(T_r)$ are finite in the absence of resetting, but in contrast to (c) here we have $\sigma(T_{r=0})<\left<T_{r=0}\right>$ [panel (d): main figure]. In this phase, $\left<T_r\right>$ monotonically increases with $r$ [panel (d): inset].\\
\indent
To understand the behavior of the mean and standard deviation of the FPT in the limit $r\to 0$, we use \eref{eq:fpt_distribution} to calculate the moments of the FPT for the underlying process without resetting. The first moment $\left<T\right>\coloneqq \int_0^{\infty}tf_T(t)dt$, i.e., the mean FPT, is found to diverge for weakly repulsive/attractive logarithmic potential, where $-1<\beta U_0<1$. Clearly, the standard deviation of the FPT, $\sigma(T)\coloneqq \sqrt{\left<T^2\right>-\left<T\right>^2}$, also diverges in this regime, which supports our observation in panel (a). For $\beta U_0> 1$, using \eref{eq:fpt_distribution} we obtain the following closed form expression for the mean FPT in the absence of resetting 
\begin{align}
\left<T\right> =\frac{x_0^2}{2D(\beta U_0-1)},
    \label{eq:mfpt_r0}
\end{align}
where we recall that $\beta U_0=2\nu -1$ [see \eref{eq:nu_def}]. \eref{eq:mfpt_r0} shows that $\left<T\right>$ is finite for $\beta U_0>1$. The standard deviation, however, diverges whenever $\beta U_0<3$. Hence, for $1<\beta U_0<3$, the mean FPT is finite but $\sigma(T)$ diverges [panel (b)].\\
\indent
For $\beta U_0>3$, the standard deviation of the FPT for the process without resetting is finite and given by
\begin{align}
\sigma(T)= \frac{x_0^2}{\sqrt{2}D(\beta U_0-1)\sqrt{\beta U_0-3}}.
    \label{eq:stdev_r0}
\end{align}
In turn, the theory of first-passage with resetting asserts that the introduction of resetting will result in a decrease of the mean FPT whenever the ratio between the standard deviation and the mean of the FPT distribution $-$ in the absence of resetting $-$ is larger than unity, and vice versa \cite{Restart-Biophysics1,ReuveniPRL}. From Eqs.~(\ref{eq:mfpt_r0}) and (\ref{eq:stdev_r0}), we see that here this ratio, commonly known as the coefficient of variation (CV), is given by 
\begin{align}
CV(T)= \frac{\sigma(T)}{\left<T\right>}=\sqrt{\frac{2}{\beta U_0-3}}.
    \label{eq:cv_r0}
\end{align}
From \eref{eq:cv_r0}, it is evident that $CV(T)$ is greater than unity for $3<\beta U_0<5$, hence the introduction of resetting lowers the mean FPT in this case. In contrast, for $\beta U_0>5$, $CV(T)$ is less than unity [\eref{eq:cv_r0}]. Therefore, unlike previous cases, here the introduction of resetting hinders first-passage.\\
\indent
In addition to the analytical results of Eqs.~(\ref{eq:mfpt}) and (\ref{eq:stdev}), in \fref{Fig7} we also plot data for $\left<T_r\right>$ and $\sigma(T_r)$ that are obtained by Langevin dynamics simulations. These data are in good agreement with theory. The details of the numerical simulation are given in Appendix C. 
\begin{figure}[t!]
\begin{centering}
\includegraphics[width=8.1cm]{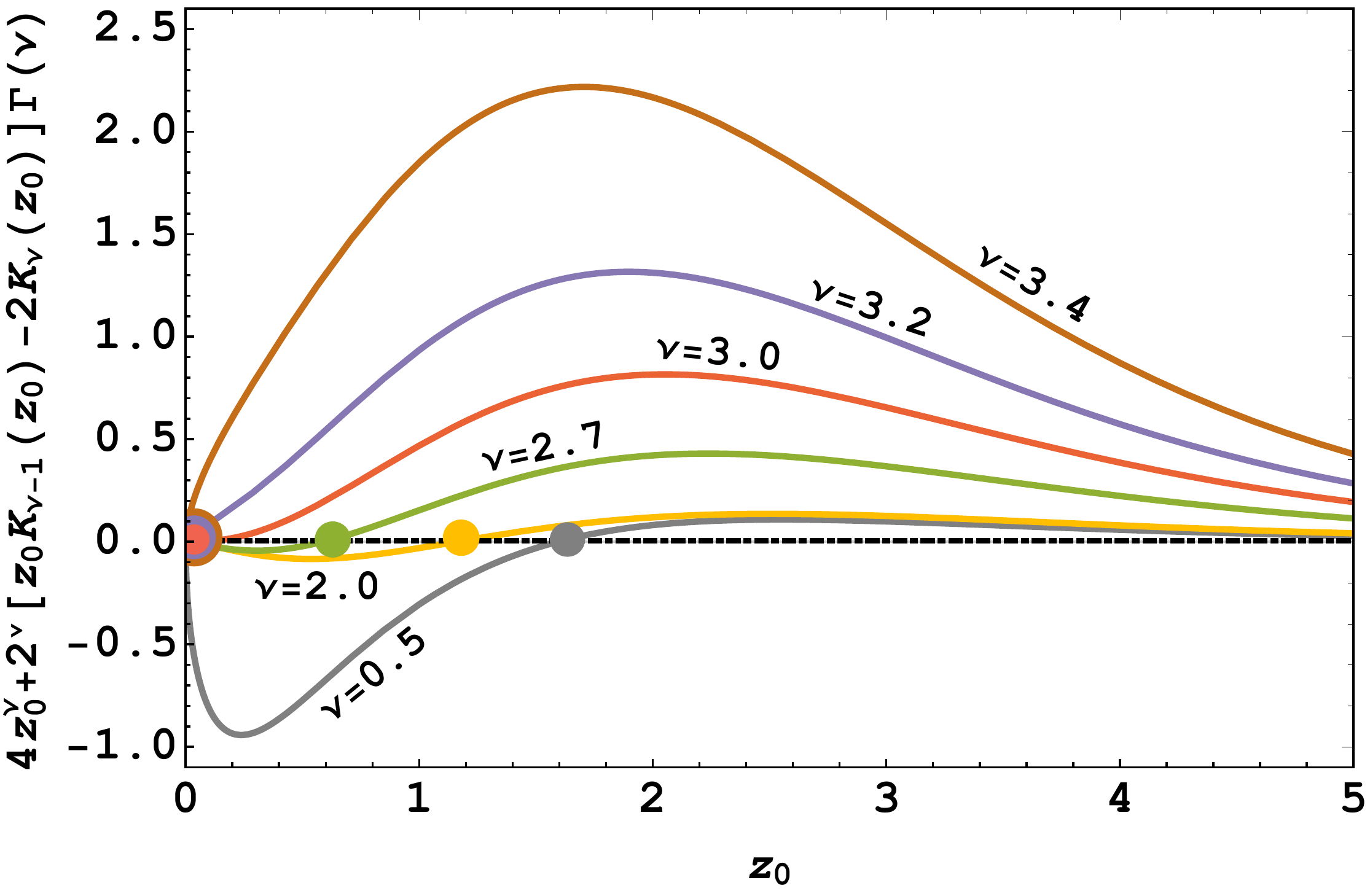}
\end{centering}
\caption{Graphical solution of \eref{eq:trans_eq} for different values of $\nu$. The solutions, $z_0^{\star}$, are highlighted by circles. }
\label{Fig8}
\end{figure}
\subsection{The resetting transition}
\indent In panels (a), (b) and (c), the variation of the mean FPT,  $\left<T_r\right>$, with the resetting rate $r$ is non-monotonic. The minimum of $\left<T_r\right>$ in each of these cases is attained at an optimal resetting rate $r^{\star}>0$. In contrast, in panel (d), $\left<T_r\right>$ shows a monotonic increase with $r$. Moreover, while  $\left<T_{r}\right>$ and $\sigma(T_{r})$ intersect in panels (a), (b) and (c), they do not intersect in panel (d). It  has recently been shown that for optimally restarted first-passage processes the mean is exactly equal to the standard deviation, i.e., $\left<T_{r^{\star}}\right> =\sigma(T_{r^{\star}}) $\cite{ReuveniPRL}. Therefore, the intersection points of $\left<T_{r}\right>$ and $\sigma(T_{r})$ in panel (a), (b) and (c) mark the optimal resetting rates. Summarizing, the results above prove that for weakly repulsive and weak to moderately attractive logarithmic potentials, the introduction of resetting accelerates first-passage to the origin, but when the attractive potential is sufficiently strong, resetting cannot expedite the process any more. This clearly indicates a \textit{resetting transition}\cite{RayReuveniJPhysA,exponent,Landau} as $\beta U_0$ grows beyond a critical value. In this subsection, we discuss this transition in terms of the optimal resetting rate $r^{\star}$. 

\begin{figure}[t!]
\begin{centering}
\includegraphics[width=8.2cm]{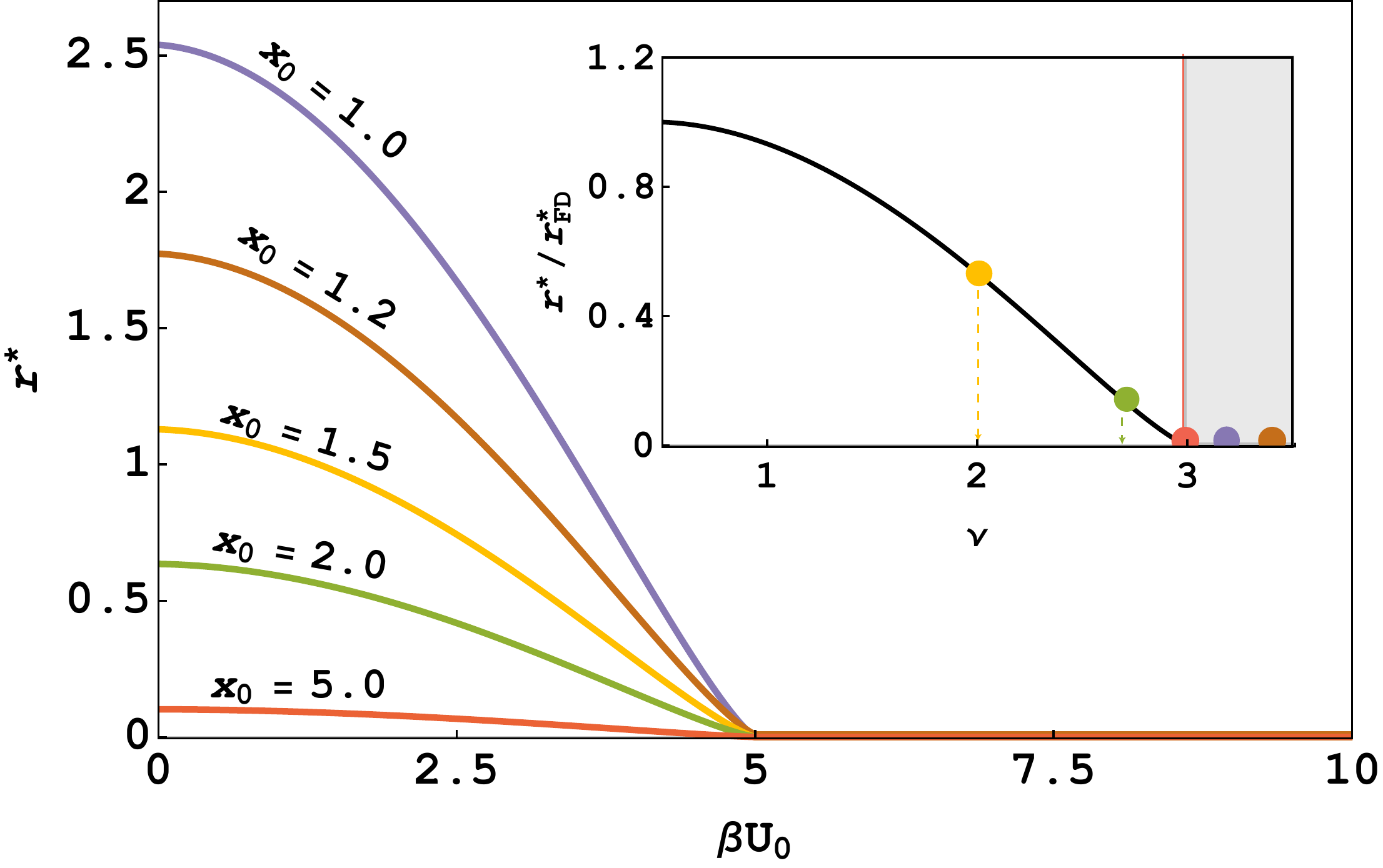}
\end{centering}
\caption{Main: The optimal resetting rate $r^{\star}$ vs $\beta U_0$ for different values of $x_0$. The resetting transition is always observed at $\beta U_0 = 5$. Beyond this point, the optimal resetting rate is zero, i.e., resetting cannot expedite first passage of the particle to the origin. Inset: The scaled optimal resetting rate $r^{\star}/r^{\star}_{FD}$ vs. the persistence exponent $\nu=(1+\beta U_0)/2$. The resetting transition is observed at $\nu=3$, shown by the vertical orange-red line. The colored circles indicate different values of $\nu$ from \fref{Fig8}.}
\label{Fig9}
\end{figure}

In order to explicitly calculate the optimal resetting rate, we recall from Sec. III that $z_0\coloneqq\alpha_0 x_0=\sqrt{rx_0^2/D}$. Therefore,
\begin{equation}
r=\frac{Dz_0^2}{x_0^2}
\label{eq:r_def}
\end{equation}
Plugging \eref{eq:r_def} into \eref{eq:mfpt}, we get the mean FPT in terms of $z_0$ as
\begin{equation}
\left<T_r\right>=\frac{x_0^2}{Dz_0^2}\left[\frac{2^{\nu-1}\Gamma{(\nu)}}{z_0^{\nu}K_{\nu}(z_0)}-1\right].\\
\label{eq:mfpt_z}
\end{equation}
When the process is reset at an optimal rate, $r=r^{\star}$, the mean FPT is minimized hence $\left[d\left<T_r\right>/dr\right]_{r=r^{\star}}=0$. From \eref{eq:r_def} we get $d\left<T_r\right>/dr\equiv\; \left( x_0^2/2z_0D\right)d\left<T_r\right>/dz_0$, which at the optimal resetting rate leads to the following transcendental equation
\begin{equation}
4z_0^{\nu}+2^{\nu}\left[z_0K_{\nu-1}(z_0)-2K_{\nu}(z_0)\right]\Gamma{(\nu)}=0.
\label{eq:trans_eq}
\end{equation}
In \fref{Fig8}, we graphically solve \eref{eq:trans_eq} for different values of $\nu$. The solutions, denoted $z_0^{\star}$, can then be utilized to calculate the optimal resetting rates $r^{\star}$ following \eref{eq:r_def}. In \fref{Fig9}, we plot these optimal resetting rates vs. $\beta U_0$ for different values of the initial position $x_0$. It is evident from the plot that the point of the resetting transition is always at $\beta U_0=5$, which does not depend on $x_0$. In other words, the introduction of resetting expedites first passage only when the ratio between the strength of the potential and the thermal energy ($\beta^{-1}$) is less than a certain critical value, $\beta U_0 < 5$. This critical value is universal in the sense that it is not affected by the initial distance of the particle from the absorbing boundary. This supports the discussion following  \eref{eq:cv_r0} which marks the transition at $CV(T)=1$. In the inset of \fref{Fig9}, we plot the optimal resetting rate scaled by $r_{FD}^{\star}$, the optimal resetting rate for free diffusion ($U_0=0$). It can be seen that after this scaling, the optimal resetting rate is uniquely determined by the persistence exponent $\nu=(1+\beta U_0)/2$. 
\begin{figure*}[t]
\begin{centering}
\includegraphics[width=16.8cm]{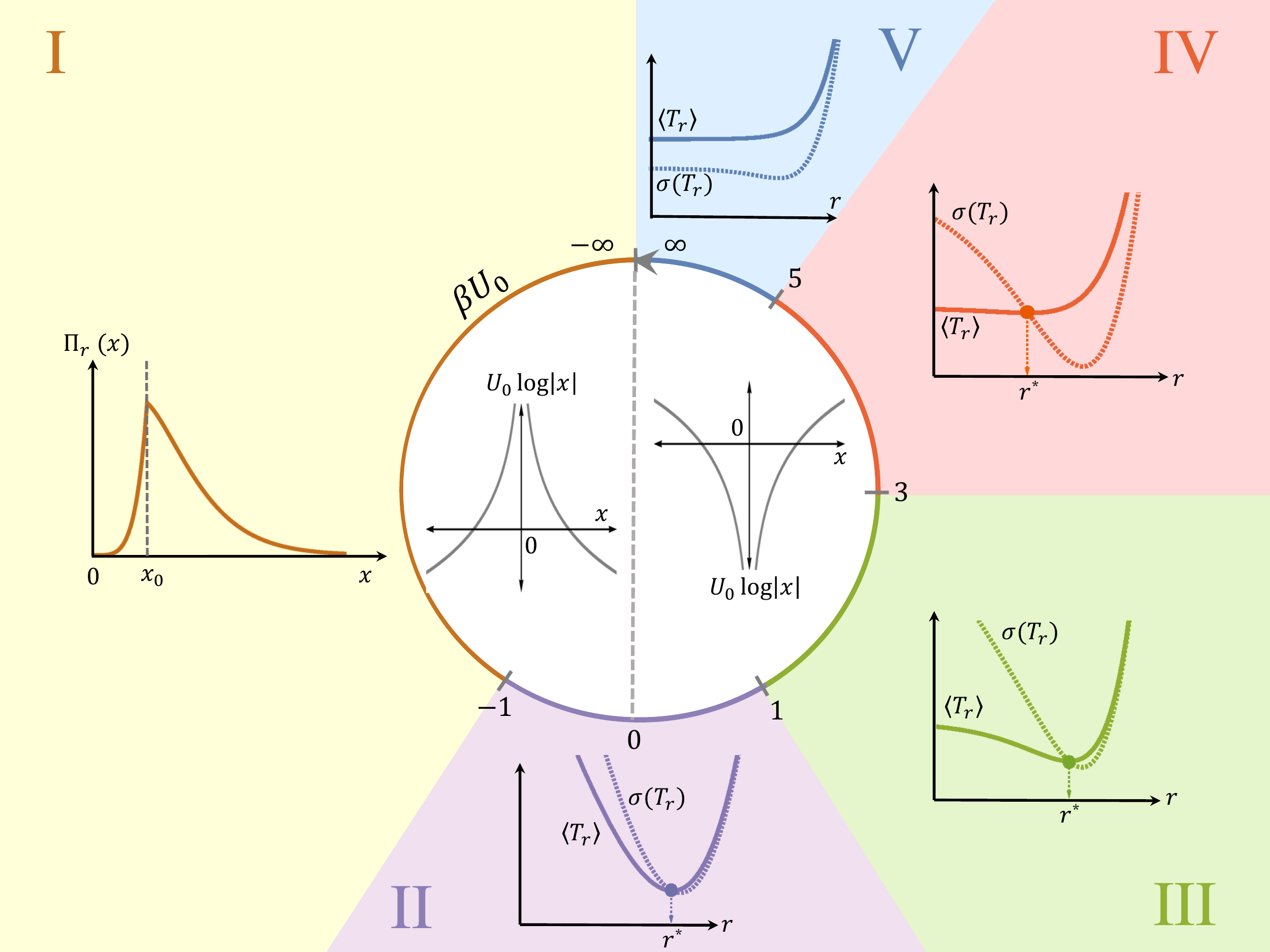}
\end{centering}
\caption{A phase diagram illustrating the possible effects of stochastic resetting on diffusion in a logarithmic potential $U(x) = U_0\log|x|$. The  potential is repulsive for $U_0<0$ and attractive for $U_0>0$, as shown in the inner circle. Phase transitions occur as $\beta U_0$, the ratio between the potential strength and the thermal energy, is tuned. Phase I: A particle, starting from $x_0$, will never reach the origin. Resetting the particle to $x_0$ at a rate $r$ then results in a nonequilibrium steady state. Phases II-V: A particle, starting from $x_0$, will eventually reach the origin. However, in the different phases the behaviour of the mean,  $\left<T_r\right>$, and standard deviation, $\sigma(T_r)$, of the first-passage time is markedly different in the limit $r \to 0$ as elaborately discussed in the main text. In particular, while in phases II-IV the introduction of resetting lowers the mean first passage time, i.e., expedites first passage to the origin, in phase V it is the other way around. The transition between these two markedly different behaviours happens at $\beta U_0=5$ irrespective of the particle's initial position.}
\label{Fig10}
\end{figure*}

\section{Conclusions}
In this work, we presented a comprehensive analysis of the effect of stochastic resetting on diffusion in a logarithmic potential $U(x)=U_0\log{|x|}$. Here $U_0$ denotes the strength of the potential which is attractive for $U_0>0$ and repulsive for $U_0<0$. We found that the effect of resetting on the dynamics of a particle that diffuses in this potential is guided solely by the interplay between this attraction/repulsion energy, $U_0$, and the thermal energy $\beta^{-1}$. This allows us to construct a phase diagram where transitions between phases occur as the dimensionless parameter $\beta U_0$ is tuned. \\
\indent
In \fref{Fig10}, we show that the entire range $\beta U_0 \in \{-\infty,\infty\}$ can be divided into five distinct phases. For $-\infty<\beta U_0<-1$, i.e., for a strongly repulsive potential, the diffusing particle can never reach the origin. In other words, the total probability of finding the particle in the ray $(0,\infty)$ [assuming it started its motion at $x_0>0$] does not decay with time. The introduction of resetting in this case leads to a steady state which marks phase I in \fref{Fig10}. Detailed analysis of this steady state was provided in Section III. \\
\indent
For $\beta U_0>-1$, i.e., when the potential is either weakly repulsive or attractive, the particle is assured to eventually hit the absorbing boundary at the origin. Four phases can then be told apart. For $-1<\beta U_0<1$ (weakly repulsive/attractive potential) both the mean and the standard deviation of the FPT diverge in absence of resetting, which marks phase II. For $1<\beta U_0<3$, $\left<T_{r= 0}\right>$ is finite, but the standard deviation diverges as $r\to 0$, marking phase III. For $\beta U_0>3$, both the mean and the standard deviation of the FPT in absence of resetting are finite. However, while for $3<\beta U_0<5$, the standard deviation is greater than the mean (phase IV), for $5<\beta U_0<\infty$ it is the other way around (phase V). \\
\indent
Following the general theory of first-passage with resetting \cite{ReuveniPRL}, we can summarize the observations as follows. For phases II, III and IV, i.e., for weakly repulsive or weak to moderately attractive potentials, introduction of resetting accelerates the first-passage of the particle to the origin. In these phases, the $\left<T_r\right>$ and $\sigma(T_r)$ curves intersect each other at an optimal resetting rate $r^{\star}$, which minimizes the MFPT. In marked contrast, for phase V, i.e., for a strongly attractive potential, resetting only delays  first-passage to the origin. In this phase, the $\left<T_r\right>$ and $\sigma(T_r)$ curves do not intersect. \\
\indent
The effect of resetting on the first-passage of a physical system can, in principle, be inverted by varying the system parameters, e.g. temperature, concentration etc., across some critical point, which gives rise to a ``resetting transition'' \cite{RayReuveniJPhysA,exponent,Landau}. For diffusion in a  potential $U(x)=U_0\log|x|$, we showed that the first-passage to the origin can be accelerated by resetting only when $\beta U_0 < 5$. This marks the point of the resetting transition at $\beta U_0=5$, which is the boundary between phase IV and phase V. For the current problem, this transition point is universal in the sense that it is independent of the initial distance of the particle from the absorbing boundary.
\noindent
\section*{ACKNOWLEDGEMENTS}
\noindent
S. Ray acknowledges support from the Raymond and Beverly Sackler Center for Computational Molecular and Materials Science at Tel Aviv University. S. Reuveni acknowledges support from the Azrieli Foundation, from the Raymond and Beverly Sackler Center for Computational Molecular and Materials Science at Tel Aviv University, and from the Israel Science Foundation (grant No. 394/19). S. Ray is thankful to Arnab Pal for insightful discussions. The authors thank Arnab Pal and Sarah Kostinski for reading and commenting on early versions of this manuscript. 
\section*{Appendix A:  Steady state density by solving \eref{eq:fper_ss}}
\renewcommand{\theequation}{A.\arabic{equation}}
Eq.~(\ref{eq:fper_ss}) in the main text is a second-order, linear and inhomogeneous differential equation. Here, we solve \eref{eq:fper_ss} for $\Pi_r(x)$, the steady state density for diffusion under resetting in a strongly repulsive logarithmic potential ($\beta U_0<-1$, i.e., $\nu<0$ [see \eref{eq:nu_def}]). To do that, we perform the following variable transformation
\begin{equation}
\Pi_r(x) \equiv x^{\rho}y(x).
    \label{eq:var_transform}
\end{equation}
Plugging \eref{eq:var_transform} into \eref{eq:fper_ss} and setting $\alpha_0=\sqrt{r/D}$ we obtain
\begin{eqnarray}
\dfrac{d^2 y(x)}{d x^{2}}+
c_1\left(\frac{1}{x}\right)\dfrac{d y(x)}{d x} 
+\left[\frac{c_2}{x^2}-\alpha_0^2\right]y(x)= 
-\alpha_0^2\frac{\delta(x-x_0)}{x^{\rho}},\nonumber\\
    \label{eq:fper_ss_y}
\end{eqnarray}
where $c_1=(2\rho+\beta U_0$), and $c2 = (\rho-1)(\rho+\beta U_0)$. In Eqs. (\ref{eq:var_transform}) and (\ref{eq:fper_ss_y}), $\rho$ is an arbitrary parameter. Setting $c_1=1$, we get $\rho=(1-\beta U_0)/2\equiv (1-\nu)$, which gives $c_2=-[(1+\beta U_0)/2]^2\equiv-\nu^2$. Therefore, \eref{eq:fper_ss_y} reduces to
\begin{eqnarray}
\dfrac{d^2 y(x)}{d x^{2}}+
\left(\frac{1}{x}\right)\dfrac{d y(x)}{d x} 
-\left[\frac{\nu^2}{x^2}+\alpha_0^2\right]y(x)= 
-\alpha_0^2x^{\nu-1}\delta(x-x_0).\nonumber\\
    \label{eq:y_bessel_equation}
\end{eqnarray}
\eref{eq:y_bessel_equation} is an inhomogeneous modified Bessel equation; its general solution\cite{BesselSol} is given by
\begin{eqnarray}
y(x)=
\begin{cases}
  A_1 I_{-\nu}(\alpha_0 x)+B_1 K_{-\nu}(\alpha_0 x) \;\;\mbox{for}\;\; x\geq x_0, \\ 
  A_2 I_{-\nu}(\alpha_0 x)+B_2 K_{-\nu}(\alpha_0 x)  \;\;\mbox{for}\;\; x< x_0,
\end{cases}
    \label{eq:y_bessel_solution}
\end{eqnarray}
where $I_{-\nu}(\cdot)$ and $K_{-\nu}(\cdot)$ are modified Bessel functions of the first and second kind with order $\nu<0$. Since $\Pi_r(x)=x^{1-\nu}y(x)$, the general solution of \eref{eq:fper_ss} reads
\begin{eqnarray}
\Pi^+_r(x) = 
  A_1x^{1-\nu}I_{-\nu}(\alpha_0 x)+B_1x^{1-\nu}K_{-\nu}(\alpha_0 x) \;\;\mbox{for}\;\; x\geq x_0,\nonumber\\ 
  \Pi^-_r(x) = 
  A_2x^{1-\nu}I_{-\nu}(\alpha_0 x)+B_2x^{1-\nu}K_{-\nu}(\alpha_0 x)  \;\;\mbox{for}\;\; x< x_0.\nonumber\\ 
\label{eq:bessel_solution}
\end{eqnarray}
where $\Pi^+_r(x)$ and $\Pi^-_r(x)$ denote the left and right branches of $\Pi_r(x)$, respectively. In order to find out the specific solution of \eref{eq:bessel_solution}, we need to calculate the coefficients $A_1$, $B_1$, $A_2$ and $B_2$ explicitly, which we accomplish in the following way.\\
\indent
The steady state density $\Pi_r(x)$ must not diverge at $x\to \infty$, and that sets $A_1 = 0$.\\
\indent
In addition, $\Pi_r(x)$ must be continuous at $x=x_0$, i.e., $\Pi_r^+(x_0)=\Pi_r^-(x_0)$. This leads to
\begin{eqnarray}
 B_1= B_2+A_2\left[\frac{I_{-\nu}(\alpha_0 x_0)}{K_{-\nu}(\alpha_0 x_0)}\right].
\label{eq:continuity}
\end{eqnarray}
\indent
The third condition comes from integrating \eref{eq:fper_ss} over the narrow spatial interval $[x_0-\Delta, x_0+\Delta]$, where $|\Delta|/x_0\ll1$, which gives
\begin{eqnarray}
\left[\frac{d\Pi_r(x)}{dx}\right]_{x_0-\Delta}^{x_0+\Delta}+\beta U_0\left[\frac{\Pi_r(x)}{x}\right]_{x_0-\Delta}^{x_0+\Delta} &-&\alpha_0^2\int_{x_0-\Delta}^{x_0+\Delta}\Pi_r(x)dx \nonumber\\
&=&-\alpha_0^2.
\label{eq:derivative_jump1}
\end{eqnarray}
Here we utilize the identity $\int_{x_0-\Delta}^{x_0+\Delta}\delta(x-x_0)=1$. The condition for continuity at $x=x_0$ leads to $\lim_{\Delta\to 0}\left[\Pi_r(x)/x\right]_{x_0-\Delta}^{x_0+\Delta}=0$ and $\lim_{\Delta\to 0}\int_{x_0-\Delta}^{x_0+\Delta}\Pi_r(x)dx=0$. Therefore, \eref{eq:derivative_jump1} boils down to 
\begin{eqnarray}
\lim_{\Delta\to 0}\left[\frac{d\Pi_r(x)}{dx}\right]_{x_0-\Delta}^{x_0+\Delta}\equiv \left[\frac{d\Pi_r^+(x)}{dx}-\frac{d\Pi_r^-(x)}{dx}\right]_{x\to x_0}=-\alpha_0^2.\nonumber\\
\label{eq:derivative_jump2}
\end{eqnarray}
Note that \eref{eq:derivative_jump2} is the same as \eref{eq:slope_difference} in the main text, derived in a separate context. Calculating the slopes of $\Pi_r^{\pm}(x)$ at $x=x_0$ from \eref{eq:bessel_solution} and plugging the results in \eref{eq:derivative_jump2} we get
\begin{eqnarray}
(B_2-B_1)\left[K_{-\nu}(\alpha_0 x_0)\right.&-&\left.\alpha_0 x_0 K_{-\nu-1}(\alpha_0 x_0)\right]+A_2\left[I_{-\nu}(\alpha_0 x_0)\right.\nonumber\\
&+&\left.\alpha_0 x_0 I_{-\nu-1}(\alpha_0 x_0)\right]=\alpha_0^2x_0^{\nu}.
\label{eq:derivative_jump3}
\end{eqnarray}
\indent
The fourth and final condition comes from the fact that for $\nu<0$, the diffusing particle never reaches the absorbing boundary placed at the origin. Hence, $\Pi_r(x)$ is normalized to unity, i.e., $\int_0^{\infty}\Pi_r(x)dx\equiv\int_0^{x_0}\Pi_r^-(x)dx+\int_{x_0}^{\infty}\Pi_r(x)dx=1$, which leads to
\begin{eqnarray}
B_2\int_0^{x_0} x^{1-\nu}K_{-\nu}(\alpha_0 x)+\frac{x_0^{1-\nu}}{\alpha_0}\left[A_2I_{1-\nu}(\alpha_0 x_0)\right.&+&\left.B_1K_{1-\nu}(\alpha_0 x_0)\right]
\nonumber\\
&=&1,
\label{eq:normalization}
\end{eqnarray}
since $\int_0^{x_0} x^{1-\nu}I_{-\nu}(\alpha_0 x)dx=\alpha_0^{-1} x_0^{1-\nu}I_{1-\nu}(\alpha_0 x_0)$ and $\int_{x_0}^\infty x^{1-\nu}K_{-\nu}(\alpha_0 x)dx=\alpha_0^{-1} x_0^{1-\nu}K_{1-\nu}(\alpha_0 x_0)$. Utilizing the general relation $I_{\mu}(y)K_{\mu+1}(y)+K_{\mu}(y)I_{\mu+1}(y)=1/y$ that is valid for $\mu\in\mathbb{C}$, we solve Eqs. (\ref{eq:continuity}), (\ref{eq:derivative_jump3}), and (\ref{eq:normalization}) to get the explicit expressions for $A_2$, $B_1$ and $B_2$
\begin{eqnarray}
B_1 &=& \alpha_0^2x_0^{\nu}I_{-\nu}(\alpha_0 x_0)\nonumber\\
A_2 &=& \alpha_0^2x_0^{\nu}K_{-\nu}(\alpha_0 x_0)\nonumber\\
B_2 &=& 0.
\label{eq:coefficient}
\end{eqnarray}
Plugging \eref{eq:coefficient} into \eref{eq:bessel_solution}, we get \eref{eq:steady_state} in the main text.
\section*{Appendix B: Laplace transform of \eref{eq:fpe_solution} for strongly repulsive potential}
\renewcommand{\theequation}{B.\arabic{equation}}
\eref{eq:fpe_solution} in the main text gives $p(x,t)$, the probability density of finding a particle at position $x$ at time $t$, when the particle diffuses in a logarithmic potential. Here, we calculate the Laplace transform of \eref{eq:fpe_solution} for $\beta U_0<-1$ ($\nu<0$), i.e., when the potential is strongly repulsive. Letting $\tilde{p}(x,s)\coloneqq \int_0^{\infty}e^{-st}p(x,t)dt$ denote the Laplace transform of $p(x,t)$, we see that for $\nu<0$
\begin{align}
\tilde{p}(x,s)= \frac{x}{2D}\left(\frac{x_0}{x}\right)^{\nu}\int_0^{\infty}\frac{e^{-st}}{t}\exp{\left(-\frac{x^2+x_0^2}{4Dt}\right)}I_{-\nu}\left(\frac{xx_0}{2Dt}\right)dt.
\label{eq:probability_repulsive}
\end{align}
Next, we note the identity \cite{integration_table} 
\begin{eqnarray}
&&\int_{0}^{\infty}e^{-st}\left(\frac{1}{t}\right)\exp{\left(-\frac{a+b}{2t}\right)}I_{\mu}\left(\frac{a-b}{2t}\right)dt \nonumber\\
&&=2 K_{\mu}\left(\sqrt{s}\left(\sqrt{a}+\sqrt{b}\right)\right)
I_{\mu}\left(\sqrt{s}\left(\sqrt{a}-\sqrt{b}\right)\right),\nonumber\\
\label{eq:Laplace_p}
\end{eqnarray}
which holds for $Re(a)\geq Re(b)>0$. Setting $\mu=-\nu$ in \eref{eq:Laplace_p}, we compare the integral with that in \eref{eq:probability_repulsive} to identify $(a+b)\equiv(x^2+x_0^2)/2D$ and $(a-b)\equiv(xx_0)/D$, which gives
\begin{eqnarray}
\sqrt{a}=\frac{(x+x_0)}{2\sqrt{D}}\;\;\mbox{and}\;\;
\sqrt{b}=\pm\frac{(x-x_0)}{2\sqrt{D}}.
\label{eq:ab}
\end{eqnarray}
Therefore, when $x\geq x_0$, $\sqrt{b}=(x-x_0)/2\sqrt{D}$ and when $x<x_0$, $\sqrt{b}=(x_0-x)/2\sqrt{D}$, leading to 
\begin{align}
(\sqrt{a}+\sqrt{b})=\frac{x}{\sqrt{D}},\;\;(\sqrt{a}-\sqrt{b})=\frac{x_0}{\sqrt{D}}\;\;\;\mbox{for}\;\;\;x\geq x_0\nonumber\\
(\sqrt{a}+\sqrt{b})=\frac{x_0}{\sqrt{D}},\;\;(\sqrt{a}-\sqrt{b})=\frac{x}{\sqrt{D}}\;\;\;\mbox{for}\;\;\;x<x_0
\label{eq:ab2}
\end{align}
Plugging these into \eref{eq:Laplace_p} and then substituting the integral in \eref{eq:probability_repulsive} accordingly, we get \eref{eq:fpt_laplace} in the main text by setting $s=r$.
\section*{Appendix C: Details of numeric simulations}
\renewcommand{\theequation}{C.\arabic{equation}}
\indent
The Langevin description for a particle diffusing in a potential $U(x)=U_0\log|x|$ in the overdamped limit reads
\begin{eqnarray}
\dot{x}=
-\frac{U_0}{\zeta x}+\eta(t),
\label{eq:langevin}
\end{eqnarray}
where $\zeta$ is the friction coefficient and $\eta(t)$ denotes a white Gaussian noise with zero mean, i.e., $\left<\eta(t)\right>=0$ and $\left<\eta(t)\eta(t^{\prime})\right>=2D\delta(t-t^{\prime})$. Here we numerically solve the above stochastic differential equation under resetting with a constant rate $r$. This implies that the position of the particle $x$ is reset to $x_0$ following random time intervals whose lengths are taken from an exponential distribution with mean $r^{-1}$.\\
\indent
To numerically calculate the steady state density $\Pi_r(x)$, we utilize {\it Heun's method} (commonly known as the \textit{second order Runge-Kutta method})\cite{Xavier} with step size $h=10^{-3}$ to perform the Langevin dynamics simulation and obtain the position $x$ of the particle in the long time limit for $10^6$ realizations. The steady state distribution is then estimated from the data and compared against the analytical results of \eref{eq:steady_state} in Fig. \ref{Fig1}. The theoretical results and the simulation data are in excellent agreement.\\
\indent
To numerically calculate the mean, $\left<T_r\right>$, and the standard deviation, $\sigma(T_r)$, of the FPT to the origin, we perform the Langevin dynamics simulation utilizing {\it Heun's method} with step size $h=10^{-4}$ or $h=10^{-6}$, depending on the system parameters. Smaller step size is necessary to minimize errors in the following cases. For weakly repulsive potential, i.e., $-1<\beta U_0<0$, smaller step size is required to lower down the otherwise high probability of losing the particle downhill when it hits the absorbing boundary at the origin but not immediately detected. For strongly attractive potential, $\beta U_0>5$, the first passage time $T_r$ reduces considerably in magnitude, which might lead to large errors if the step size is not reduced accordingly. The simulation is performed for $10^6$ realizations in each case to calculate the mean and the standard deviation of the FPT. The simulation data show very good agreement with the analytical results of \eref{eq:mfpt} and \eref{eq:stdev}, as shown in \fref{Fig7}.
\section*{Appendix D: First passage time by solving \eref{eq:bme_Q}}
\renewcommand{\theequation}{D.\arabic{equation}}
\eref{eq:bme_Q} in the main text is a second order, linear and inhomogeneous partial differential equation. It describes the time evolution of $Q_r(t|x_0)$, the survival probability of a particle that diffuses under resetting in a weakly repulsive or attractive logarithmic potential ($\beta U_0>-1$), in the presence of an absorbing boundary placed at the origin. Here, we solve \eref{eq:bme_Q} in the Laplace space to calculate the FPT of the particle to the absorbing boundary. To do that, we set  $\tilde{Q_r}(s|x_0)\coloneqq\int_{0}^{\infty}Q_r(t|x_0)e^{-st}dt$ as the Laplace transform of $Q_r(t|x_0)$. Laplace transforming \eref{eq:bme_Q} we get
\begin{eqnarray}
\dfrac{d^2 \tilde{Q_r}(s|x_0)}{d x_0^{2}}-\left(\frac{\beta U_0}{x_0}\right)\dfrac{d \tilde{Q_r}(s|x_0) }{d x_0}&-&\left(\frac{s+r}{D}\right)\tilde{Q_r}(s|x_0)=\nonumber\\
&-&\frac{1+r\tilde{Q_r}(s|x_r)}{D},
    \label{eq:Q_laplace}
\end{eqnarray}
where we utilized the initial condition $Q_r(0|x_0)=1$. In order to convert \eref{eq:Q_laplace} to a homogeneous differential equation, we consider a shifted observable $\tilde{q}(s|x_0)\coloneqq\tilde{Q_r}(s|x_0)-\left[(1+r\tilde{Q_r}(s|x_r))/(s+r)\right]$. \eref{eq:Q_laplace} in terms of $\tilde{q}(s|x_0)$ then reads
\begin{align}
\dfrac{d^2 \tilde{q}(s|x_0)}{d x_0^{2}}-\left(\frac{\beta U_0}{x_0}\right)\dfrac{d \tilde{q}(s|x_0) }{d x_0}-\left(\frac{s+r}{D}\right)\tilde{q}(s|x_0)=0.
    \label{eq:q_laplace}
\end{align}
Next, we consider another change of variables, viz., $\tilde{q}(s|x_0)=x_0^{\nu}\tilde{y}(s|x_0)$ with $\nu={(1+\beta U_0)/2}$. \eref{eq:q_laplace} then reduces to
\begin{align}
x_0^2\dfrac{d^2 \tilde{y}(s|x_0)}{d x_0^{2}}+x_0\dfrac{d \tilde{y}(s|x_0) }{d x_0}-\left(\alpha^2x_0^2+\nu^2\right)\tilde{y}(s|x_0)=0,
    \label{eq:y_laplace}
\end{align}
where $\alpha=\sqrt{(s+r)/D}$. \eref{eq:y_laplace} is a modified Bessel equation with the general solution\cite{BesselSol} \begin{align}
\tilde{y}(s|x_0)=A(s)I_{\nu}(\alpha x_0)+B(s)K_{\nu}(\alpha x_0),
    \label{eq:bessel_general}
\end{align}
where $I_{\nu}(\alpha x_0)$ and $K_{\nu}(\alpha x_0)$ are modified Bessel functions of the first and second kind of order $\nu>0$, respectively. To obtain the specific solution of \eref{eq:y_laplace}, we need to determine the explicit expressions of $A(s)$ and $B(s)$. We accomplish that in the following way. \\
\indent
We note that in the limit $x_0\rightarrow \infty$, i.e., when the initial position is very far from the absorbing boundary, chances are negligible that the particle is absorbed, and hence the survival probability is unity; $Q_r(t|x_0)|_{x_0\rightarrow\infty}=1$. That leads to $\tilde{Q_r}(s|\infty)=1/s$ and $\tilde{q}(s|\infty)=1/s-\left[(1+r\tilde{Q_r}(s| x_r))/(s+r)\right]$, which in turn gives $\tilde{y}(s|\infty)=0$. Since $K_{\nu}(\infty)\rightarrow 0$, but $I_{\nu}(\infty)$ diverges, we set $A(s)=0$ to keep things consistent. \\
\indent
To calculate $B(s)$, we first assume that the absorbing boundary is kept at $x=\epsilon$. This leads to $Q_r(t|x_0)|_{x_0=\epsilon}=0$, since the survival probability vanishes if the process is initiated at the absorbing boundary. Therefore,  $\tilde{Q_r}(s|\epsilon)=0$ and $\tilde{q}(s|\epsilon)=-\left[(1+r\tilde{Q_r}(s| x_r))/(s+r)\right]$. Thus by definition $\tilde{y}(s|\epsilon)=-\epsilon^{-\nu}\left[(1+r\tilde{Q_r}(s|x_r))/(s+r)\right]\equiv B(s)K_{\nu}(\alpha\epsilon)$. Taking $\epsilon\rightarrow0$ to match the conditions of our problem, we have $K_{\nu}(\alpha \epsilon)\simeq (\alpha \epsilon)^{-\nu}\left[2^{\nu-1}\Gamma(\nu)\right]$, where $\Gamma{(\nu})$ denotes the gamma function. Comparing, we get $B(s)=-\alpha^{\nu}(1+r\tilde{Q_r}(s|x_r))/\left[2^{\nu-1}\Gamma(\nu)(s+r)\right]$. Thus the specific solution to \eref{eq:y_laplace} reads
\begin{align}
\tilde{y}(s|x_0)=-\frac{1+r\tilde{Q_r}(s|x_r)}{2^{\nu-1}\Gamma(\nu)(s+r)}\alpha^{\nu}K_{\nu}(\alpha x_0).
    \label{eq:bessel_y}
\end{align}
Therefore, the specific solution to \eref{eq:Q_laplace} is
\begin{align}
\tilde{Q_r}(s|x_0)=-\frac{1+r\tilde{Q_r}(s|x_r)}{s+r}\left[\frac{(\alpha x_0)^{\nu}K_{\nu}(\alpha x_0)}{2^{\nu-1}\Gamma(\nu)}-1\right].
    \label{eq:bessel_Q}
\end{align}
We can evaluate $\tilde{Q_r}(s|x_0)$ in a self consistent manner by putting $x_r=x_0$ in the above equation, which means that the particle is taken back to the initial position after resetting. After simplification, this gives
\begin{align}
\tilde{Q_r}(s|x_0)=\frac{2^{\nu-1}\Gamma(\nu)-(\alpha x_0)^{\nu}K_{\nu}(\alpha x_0)}{s\; 2^{\nu-1}\Gamma(\nu)+r\;(\alpha x_0)^{\nu}K_{\nu}(\alpha x_0)},
    \label{eq:survival_solution}
\end{align}
which is equivalent to \eref{eq:survival_solution_main} in the main text.
Recalling that $f_{T_r}(t)=-dQ_r(t|x_0)/dt$\cite{gardiner}, where $f_{T_r}(t)$ the probability density of $T_r$, we see that \eref{eq:survival_solution} allows us to calculate any moment of $T_r$ following the relation
\begin{eqnarray}
\left<T_r^n\right>&=&-\int_{0}^{\infty}t^n\left[\frac{\p Q_r(t|x_0)}{\p t}\right]dt\nonumber\\
&=&\left(-1\right)^{n-1}n\left[\frac{d^{n-1}\tilde{Q_r}(s|x_0)}{ds^{n-1}}\right]_{s=0}.
\label{eq:fpt_moments}
\end{eqnarray}
In particular, the first moment or the mean FPT is given by $\left<T_r\right>=\left[\tilde{Q_r}(s|x_0)\right]_{s=0}$.   Indeed, setting $s=0$ in \eref{eq:survival_solution}, we recover \eref{eq:mfpt} of the main text.
\section*{Appendix E:  Laplace transform of \eref{eq:fpt_distribution}}
\renewcommand{\theequation}{E.\arabic{equation}}
\eref{eq:fpt_distribution} presents $f_T(t)$, the first passage time distribution to the origin for a particle that diffuses in a weakly repulsive or attractive logarithmic potential with $\beta U_0>-1$. Letting $\tilde{T}(s)\coloneqq\int_0^{\infty}e^{-st}f_T(t)dt$ denote the Laplace transform of $f_T(t)$, we utilize \eref{eq:fpt_distribution} to get
\begin{eqnarray}
\tilde{T}(s)=\frac{1}{\Gamma(\nu)}\left(\frac{x_0^2}{4D}\right)^{\nu}\int_0^{\infty}t^{-(\nu+1)} \exp{\left(-st-\frac{x_0^2}{4Dt}\right)}dt.\nonumber\\
\label{eq:fpt_lt_def}
\end{eqnarray}
Next, we note the following relation \cite{zwillinger}
\begin{align}
\int_0^{\infty}x^{\gamma-1} \exp{\left(-x-\frac{\mu^2}{4x}\right)}dx = 2\left(\frac{\mu}{2}\right)^{\gamma}K_{-\gamma}(\mu),
\label{eq:zwilinger}
\end{align}
which holds for $|arg (\mu)|<\pi/2$ and $Re(\mu^2)>0$. Comparing the integrals in Eqs.~(\ref{eq:zwilinger}) and (\ref{eq:fpt_lt_def}) (\ref{eq:zwilinger}), we identify $x\equiv st$, $\gamma \equiv -\nu$ and $\mu \equiv \sqrt{sx_0^2/D}$. Evaluating the integral in \eref{eq:fpt_lt_def} with the aid of \eref{eq:zwilinger} and plugging the result back into \eref{eq:fpt_lt_def}, we obtain \eref{eq:fpt_lt} in the main text.

\end{document}